\pdfoutput=1
\documentclass{article}

\usepackage[final]{neurips_2024}

\usepackage[utf8]{inputenc} 
\usepackage[T1]{fontenc}    
\usepackage[hidelinks]{hyperref} 
\usepackage{url}            
\usepackage{booktabs}       
\usepackage{amsfonts}       
\usepackage{nicefrac}       
\usepackage{microtype}      
\usepackage{xcolor}         
\usepackage[hang,flushmargin]{footmisc}
\usepackage{enumitem}
\usepackage{xspace}
\usepackage{arydshln}
\usepackage{multirow}
\usepackage{graphicx}
\usepackage{mdframed}
\usepackage{pifont}
\usepackage{tikz}
\usepackage{subcaption}
\usepackage{longtable}
\usepackage{afterpage}

\usepackage{natbib}
\usepackage{array}
\usepackage{amsmath}
\usepackage{tabularx}
\usepackage{xcolor,colortbl}
\usepackage{array}           
\usepackage{dashrule}        
\usepackage{longtable}

\usepackage{etoolbox}
\makeatletter
\patchcmd{\@verbatim}
  {\verbatim@font}
  {\verbatim@font\small}
  {}{}
\makeatother

\newcommand\ignore[1]{}

\let\oldFootnote\footnote
\newcommand\nextToken\relax

\renewcommand\footnote[1]{%
    \oldFootnote{#1}\futurelet\nextToken\isFootnote}

\newcommand\isFootnote{%
    \ifx\footnote\nextToken\textsuperscript{,}\fi}

\newcommand{\umb}{UMBRELA\xspace}

\newcommand{\autonuggets}{Auto\-Nuggets\xspace}
\newcommand{\autonuggetsedits}{Auto\-Nuggets+Edits\xspace}
\newcommand{\autoassign}{Auto\-Assign\xspace}

\newcommand{\manulassign}{Manual\-Assign\xspace}

\begin{document}

\title{Initial Nugget Evaluation Results for the TREC 2024 RAG Track with the AutoNuggetizer Framework}

\author{
Ronak Pradeep$^1$, Nandan Thakur$^1$, Shivani Upadhyay$^1$,\\
{\bf Daniel Campos$^2$, Nick Craswell$^3$, Jimmy Lin$^1$} \\[1ex]
$^1$ University of Waterloo \quad $^2$ Snowflake \quad $^3$ Microsoft
}

\pagestyle{empty}

\maketitle

\begin{abstract}
This report provides an initial look at partial results from the TREC 2024 Retrieval-Augmented Generation (RAG) Track.
We have identified RAG evaluation as a barrier to continued progress in information access (and more broadly, natural language processing and artificial intelligence), and it is our hope that we can contribute to tackling the many challenges in this space.
The central hypothesis we explore in this work is that the nugget evaluation methodology, originally developed for the TREC Question Answering Track in 2003, provides a solid foundation for evaluating RAG systems.
As such, our efforts have focused on ``refactoring'' this methodology, specifically applying large language models to both {\it automatically} create nuggets and to {\it automatically} assign nuggets to system answers.
We call this the AutoNuggetizer framework.
Within the TREC setup, we are able to calibrate our fully automatic process against a manual process whereby nuggets are created by human assessors semi-manually and then assigned manually to system answers.
Based on initial results across 21 topics from 45 runs, we observe a strong correlation between scores derived from a fully automatic nugget evaluation and a (mostly) manual nugget evaluation by human assessors.
This suggests that our fully automatic evaluation process can be used to guide future iterations of RAG systems.
\end{abstract}

\section{Introduction}

This report examines the partial results from the TREC 2024 Retrieval-Augment Generation (RAG) Track.
There is, obviously, tremendous excitement and interest in RAG, but we feel that the {\it evaluation} of RAG output remains deficient from many perspectives.
Furthermore, the lack of standardized evaluations presents a barrier to continued progress in information access, and more broadly, natural language processing and artificial intelligence.
We hope to harness the institutional knowledge and resources provided by the National Institute of Standards and Technology (NIST) via the Text Retrieval Conference (TREC) to tackle these challenges.
Now in its 33\textsuperscript{rd} year, TREC has led the way in many aspects of evaluation in information retrieval (IR), natural language processing (NLP), and beyond, producing many innovations that the community (both researchers as well as practitioners) take for granted today.
We hope that these efforts will lay similar foundations for standardization of RAG evaluation in the future.

Our central hypothesis is that the nugget evaluation methodology \citep{Voorhees_TREC2003} provides a solid foundation for evaluating RAG systems.
This methodology was originally articulated more than two decades ago in the context of the TREC Question Answering (QA) Track for evaluating answers to free-form ``definition questions''.
This matches our RAG setting, where for a given query, there are claims or facts (i.e., nuggets) from different documents that are combined into a fluent and cohesive natural language response.

Given this starting point, our collective efforts have focused on ``refactoring'' the original nugget evaluation methodology in light of large language models (LLMs).
Specifically, we leverage LLMs to both {\it automatically} create nuggets and {\it automatically} assign nuggets to system-generated answers.
We implement our nugget evaluation methodology in the AutoNuggetizer framework.
While this is not the first attempt to automate nugget evaluations~\citep{Lin_Demner-Fushman_HLT-EMNLP2005,Lin_Demner-Fushman_IR2006}, the advent of LLMs provides opportunities that did not exist before.

The biggest advantage of our approach is minimal reinvention of the wheel, in that we are able to leverage the existing body of work that has gone into exploring the nugget evaluation methodology, e.g.,~\citep{Lin_Zhang_SIGIR2007,Dang_Lin_ACL2007}.
For aspects that are not directly impacted by our use of LLMs, we can continue to assert findings from the literature without needing to carefully revalidate those claims again.

Furthermore, and unique to the TREC setup, we calibrate our fully automatic process against a (mostly) manual process whereby nuggets are created by NIST assessors semi-manually and assigned manually to system-generated answers by the same assessors.

It is important to qualify that this report focuses {\it only} on answer content (in terms of nuggets), and does not take into account {\it support}, or the requirement that the content be grounded by references into the underlying document collection.
Thus, we do not consider possible LLM hallucinations at all.
We will examine support in a separate report.

To provide the tl;dr ---

\begin{quote}
Based on initial results across 21 topics from 45 runs, we observe a strong correlation between scores derived from a fully automatic nugget evaluation and a (mostly) manual nugget evaluation by NIST assessors.
This suggests that our fully automatic evaluation process can be used to guide future iterations of RAG systems.
\end{quote}

This report attempts to provide the experimental results to support the above assertion.\footnote{We apologize in advance to our many colleagues who have done excellent work on RAG evaluation. Given the pressures of the strict TREC timeline, we did not have the opportunity to survey the field and acknowledge previous work in this report. Instead, we have decided to focus exclusively on our nugget evaluation methodology. A thorough discussion of the advantages and disadvantages of alternative approaches will be provided in a subsequent follow-up.}

\begin{table}[t]
\setlength{\tabcolsep}{3pt}

\renewcommand{\arraystretch}{1}
\centering
\resizebox{0.95\textwidth}{!}{
{\begin{tabular}{p{16cm}}
\toprule
{\bf Query}: how did african rulers contribute to the triangle trade \\[1ex]
{\bf Answer}: African rulers played a significant role in the triangular trade by capturing and supplying slaves to European traders. They actively participated in the trade by capturing individuals from inland areas and transporting them to the coast, where they were sold to European traders. This involvement was driven by the economic benefits they received, such as gold, guns, and other goods from Europe. Many African rulers and merchants developed new trade routes to avoid those who refused to cooperate, ensuring a steady supply of captives for the transatlantic slave trade. The trade was highly profitable for African rulers, who received a variety of European goods in exchange for slaves. This economic incentive led to the expansion of slave trade networks and the growth of states based on slavery, such as the Kingdom of Dahomey, which became one of the most prosperous nations due to its involvement in the slave trade. The triangular trade system involved three stages: the export of manufactured goods from Europe to Africa, the transportation of enslaved Africans to the Americas, and the return of goods such as sugar, coffee, and tobacco to Europe. African rulers were crucial in the first stage, where they exchanged captives for European goods, which were often luxury items or weapons that helped them gain military and political advantages over their rivals. Despite some African rulers voicing opposition to the practice, the lure of profits ensured their continued participation in the trade. The involvement of African rulers was essential for the success of the transatlantic slave trade, as European traders relied on them to capture and supply slaves from the African interior. This collaboration was fundamental to the operation of the triangular trade, as European traders were generally confined to coastal areas and depended on African intermediaries to procure slaves. In summary, African rulers contributed to the triangular trade by capturing and supplying slaves to European traders, driven by the economic benefits they received in exchange. Their involvement was crucial for the success and expansion of the transatlantic slave trade. \\
\bottomrule
\end{tabular}}}
\vspace{0.25cm}
\caption{Sample answer from the TREC 2024 RAG Track for topic \texttt{2024-35227}. Note that here we have purposely omitted citations in the answer since this report does not cover {\it support}.}
\label{tab:sample-sample}
\end{table}

\section{The Setup of the TREC 2024 RAG Track}

The TREC 2024 Retrieval-Augmented Generation (RAG) Track comprises three distinct but inter-connected tasks:\ Retrieval (R), Augmented Generation (AG), and full Retrieval-Augmented Generation (RAG).
Participants were given 301 queries (called topics in TREC parlance); 
their ultimate task was to return, for the AG and RAG tasks, well-formed answers for each individual query (up to a maximum of 400 words).
The Retrieval (R) task can be viewed as an intermediate product in a full RAG pipeline.

Throughout this paper, we use topic \texttt{2024-35227} ``how did african rulers contribute to the triangle trade'' as a running example.
A system-generated answer is provided in Table~\ref{tab:sample-sample}; in this case, the answer was generated using GPT-4o~\citep{ragnarok}.
Note that we have purposely omitted citations from this answer.
Actual submissions to the TREC 2024 RAG Track took the form of structured JSON data wherein each answer sentence is explicitly linked to citations of documents within the corpus that (purportedly) support the assertions made in the sentence.
This report does {\it not} consider the evaluation of support, i.e., attempts to measure the appropriateness of citations.
We will discuss this important aspect of RAG evaluation in a separate paper.

\begin{table}[t]
\setlength{\tabcolsep}{3pt}

\renewcommand{\arraystretch}{1}
\centering
\resizebox{0.95\textwidth}{!}{
{\begin{tabular}{p{16cm}}
\toprule
{\bf docid:} \texttt{msmarco\_v2.1\_doc\_27\_13195298\#7\_19215443} \\
{\bf Title:} - \\
{\bf Text:} How did some African rulers participate in the slave trade? African rulers had a major role in the slave trade. They benefited from the slave trade by capturing control of the slave trade. [...] Traders then returned to Europe with sugar, coffee, and tobacco. How did Africans change the Americas?" \\
{\bf NIST Judgment:} 3 \\
\midrule
{\bf docid:} \texttt{msmarco\_v2.1\_doc\_53\_75729873\#13\_135844381} \\
{\bf Title:} - \\
{\bf Text:} Lured by its profits, many African rulers continued to participate. African merchants developed new trade routes to avoid rulers who refused to cooperate. [...] The Africans were then transported across the Atlantic and sold in the West Indies. Merchants bought sugar, coffee, and tobacco in the West Indies and sailed to Europe with these products. \\
{\bf NIST Judgment:} 0 \\
\midrule
{\bf docid:} \texttt{msmarco\_v2.1\_doc\_37\_390360760\#3\_822422101} \\
{\bf Title:} Atlantic slave trade \\
{\bf Text:} Research published in 2006 reports the earliest known presence of African slaves in the New World. A burial ground in Campeche, Mexico, suggests slaves had been brought there not long after Hernán Cortés completed the subjugation of Aztec and Mayan Mexico. [...] The third and final part of the triangle was the return of goods to Europe from the Americas. \\
{\bf NIST Judgment:} 2 \\
\midrule
{\bf docid:} \texttt{msmarco\_v2.1\_doc\_23\_1401225076\#4\_3089103831} \\
{\bf Title:} The Triangular Trade: APUSH Topics to Study [...] \\
{\bf Text:} Twelve million Africans were captured in Africa with the intent to enter them into the slave trade. [...] The continent fell far behind the growth of the developing world, opening it up to European colonization in the 19th century. What are some historical people related to the Triangular Trade? Sir John Hawkins:\\
{\bf NIST Judgment:} 2\\
\midrule
{\bf docid:} \texttt{msmarco\_v2.1\_doc\_33\_1468082722\#2\_3121913532} \\
{\bf Title:} Roles played by leaders of African societies in continuing the trade [...] \\
At that time, there was no concept of being African – identity and loyalty were based on kinship or membership of a specific kingdom or society, rather than to the African continent. States based on slavery grew in power and influence. [...] Some African slave sellers became extremely wealthy from the expansion of the slave trade networks. \\
{\bf NIST Judgment:} 2 \\
\bottomrule
\end{tabular}}}
\vspace{0.25cm}
\caption{Sample retrieval results for topic \texttt{2024-35227} ``how did african rulers contribute to the triangle trade''.}
\label{tab:sample-retrieval}
\end{table}

The Retrieval (R) task adopts a standard {\it ad hoc} retrieval setup, where systems are tasked with returning ranked lists of relevant passages from the MS MARCO V2.1 deduped segment collection~\citep{ragnarok}, which traces its lineage back to the original MS MARCO dataset~\citep{Bajaj:1611.09268:2018}.
Unlike previous MS MARCO data, where passages were selected to be short human-readable answers, the V2.1 passages adopted a sliding window approach over full documents to generate more ``natural'' content.
These passages form segments (sometimes called ``chunks'') suitable for RAG.
Table~\ref{tab:sample-retrieval} shows an example of retrieval results (i.e., from the R task) from the MS MARCO V2.1 segment collection.
These results can then serve as prompt input for LLM-based answer generation.

In the Augmented Generation (AG) task, participants receive a fixed list of 100 retrieved results to generate their answers, provided by the organizers.
In the retrieval-augmented generation (RAG) task, participants were free to retrieve and choose any document from the corpus to generate their answers.

This paper focuses on analyzing the results of system-generated answers only from the AG and RAG tasks.
For interested readers, we have written a separate report focused specifically on the retrieval task~\citep{Upadhyay:2411.08275:2024}.

\paragraph{Corpus Characteristics.}
The evaluation corpus---MS MARCO V2.1 deduped segment collection~\citep{ragnarok}---encompasses a total of 113,520,750 text passages, i.e., segments.
These passages were derived through a data processing pipeline applied to the MS MARCO V2 document collection~\citep{craswell2022overview}.
The preparation involved two key steps:

\begin{enumerate}
    \item Document deduplication: The original collection of 11,959,635 documents underwent near-duplicate removal using locality-sensitive hashing (LSH), implementing MinHash with 9-gram shingles. This process resulted in 10,960,555 unique documents.
    \item Passage generation: The deduped document collection was segmented into text passages using an overlapping sliding window approach, incorporating the parameters presented below.
    The resulting passages were typically 500--1000 characters in length.
    \begin{itemize}
        \item A window size of 10 sentences long
        \item A stride length of 5 sentences long
    \end{itemize}
\end{enumerate}

\noindent While these segments (passages) constitute the fundamental retrieval units, we adopt the convention of referring to them as ``documents'' throughout this report.

\paragraph{Topic Selection and Characteristics.}
Our evaluation uses contemporary topics sourced from Bing Search logs, specifically selecting non-factoid queries that demand comprehensive, multi-faceted, and potentially subjective responses~\citep{Rosset:2402.17896:2024}. 
Unlike previous MS MARCO queries, which were designed to be answered in a sentence or two, this query selection method gives us queries that need to be answered by multiple nuggets, potentially extracted from multiple documents. 

The set of test topics was constructed near the submission deadline (late July 2024), rather than reusing queries that had already been publicly released in~\citet{Rosset:2402.17896:2024}. 
This timing strategy primarily addresses concerns about relevance judgment contamination, though it's worth noting that the passage corpus is likely to be already contained in LLM pretraining data given its web-based nature and open-source availability.

The final topics, curated by NIST annotators, comprise 301 topics (queries) in total.
Figure~\ref{fig:sample-topics} presents representative examples that accurately highlight the nature of real-world user information needs, including their original formulation with grammatical inconsistencies and spelling errors.

\begin{figure}[t]
    \begin{itemize}
        \item why are cancer rates higher on the east coast
        \item how using maps can impact your pedagogy
        \item why was dame van winkle portrayed so negatively
        \item what is scientific evidence for or against the use of yoghurt
        \item what target stors's policies for shoplifting
    \end{itemize}
    \caption{Examples of five topics taken from the TREC 2024 RAG Track, demonstrating the real-world user queries.}
    \label{fig:sample-topics}
\end{figure}

\paragraph{Reference Ranked Lists.}
We provided participants with reference ranked lists for those who did not wish to perform retrieval on the document collection themselves.
This represents input to the AG task, where participants could focus on the ``augmented generation'' part of RAG.
We provided the top documents for each query based on an effective multi-stage ranking pipeline, described below:

The ranked lists that we provided to participants for AG integrate a combination of both first-stage retrievers and rerankers.
For first-stage retrieval, we employed a combination of a traditional retriever, BM25 + Rocchio, available in Anserini~\citep{anserini}; GTE-L, a dense retrieval model from Alibaba-NLP~\citep{li2023towards}; and two Snowflake ArcticEmbed dense retrieval models~\citep{merrick:2024}, ArcticEmbed-M and ArcticEmbed-L.
For each method, we retrieved the top 3K documents. 
To leverage the strengths of multiple retrieval strategies, we performed reciprocal rank fusion (RRF) across all retrieved results; hybrid approaches have demonstrated improved retrieval effectiveness across multiple TREC tracks~\citep{unoduolisto}.

Following retrieval, we deployed a cascading neural reranking pipeline to further refine the ranked lists.
We used monoT5-3B~\citep{nogueira-etal-2020-document,EMD}, a pointwise reranker model that reranks the top 3K documents from the retrieval stage, which are again fused with RRF from the retrieval stage.

We finally incorporated RankZephyr, a state-of-the-art listwise reranker~\citep{pradeep:2023b}, which takes a query and a list of passages as input and produces a reordered list based on the query--document relevance.
We used RankZephyr to rerank the top 100 candidates from the retrieval + rerank stage, which we yet again fused with the results from the retrieval + rerank stage.
RankZephyr is available through the \texttt{rank\_llm} Python package, making our entire pipeline reproducible.
Finally, the top 100 reranked documents from the official collection are provided to the AG participants.

\section{The Nugget Evaluation Methodology}

\begin{figure}[t]
    \centering
    \includegraphics[width=\linewidth]{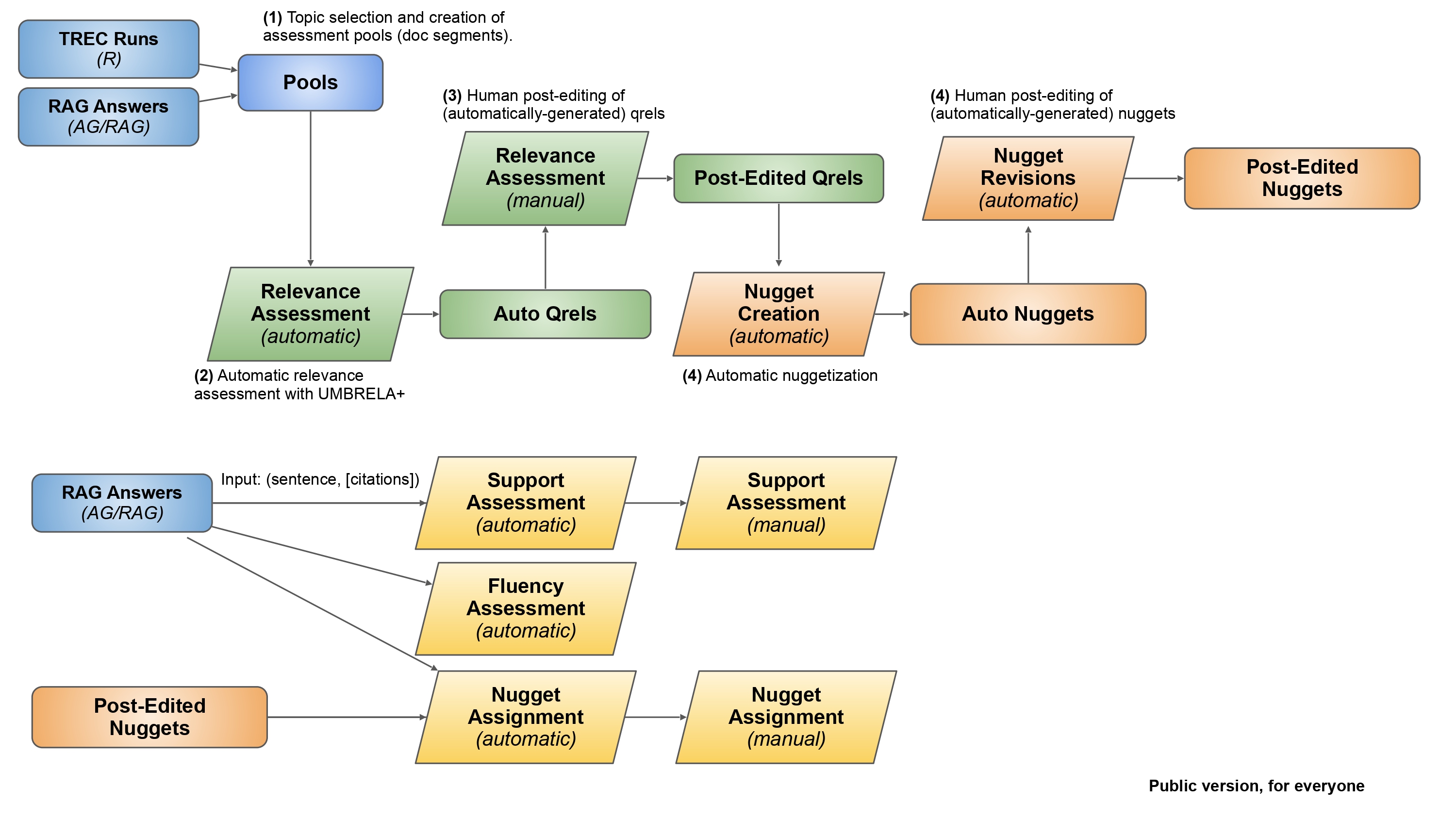}
    \caption{The high-level overview of the TREC 2024 RAG Track.}
    \label{fig:flow}
\end{figure}

A high-level overview of the evaluation flow for the TREC 2024 RAG Track is shown in Figure~\ref{fig:flow}.
In this report, we are only concerned with the nugget evaluation methodology (the orange boxes in the figure).
In particular, we do not discuss the evaluation of support (i.e., citations) at all.
For retrieval, we can point interested readers to a separate report detailing our findings~\citep{Upadhyay:2411.08275:2024}.
Note that the figure is taken from track guidelines posted in Spring 2024 and thus slightly outdated.
The actual evaluation workflow we implemented is substantively similar, but with slightly different terminology.

The gist of the nugget evaluation methodology is as follows, where we quote from \citet{Voorhees_TREC2003}:

\begin{quote}
The assessor first created a list of ``information nuggets'' about the $[$query$]$. 
An information nugget was defined as a fact for which the assessor could make a binary decision as to whether a response contained the nugget.
At the end of this step, the assessor decided which nuggets were vital---nuggets that must appear in $[$a response$]$ for that $[$response$]$ to be good. \\

The assessor went on to the second step once the nugget list was created. 
In this step the assessor went through each of the system responses in turn and marked (whether) each nugget appeared in the response.
\end{quote}

\begin{table}[t]
\setlength{\tabcolsep}{3pt}

\renewcommand{\arraystretch}{1}
\centering
\resizebox{0.95\textwidth}{!}{
{\begin{tabular}{p{16cm}}
\toprule
        {\bf Automatic nugget creation (\autonuggets) using \umb qrels} \\
\midrule
            African rulers captured and sold slaves to Europeans (vital) \\
            African rulers waged wars to capture more slaves (vital) \\
            African rulers exchanged slaves for firearms (vital) \\
            African rulers' involvement was crucial for the trade's scale (vital) \\
            African rulers' involvement increased due to European demand (vital) \\
            African rulers captured slaves from raids and wars (vital) \\
            African rulers' control and supply of captives was significant (vital) \\
            African rulers transported captives to coastal slave forts (vital) \\
            African rulers traded slaves for textiles and ironware (vital) \\
            African rulers grew wealthy from the slave trade (okay) \\
            African rulers' alliances with Europeans were fundamental (okay) \\
            African rulers' trade caused increased tension and violence (okay) \\
            African rulers' actions led to internal African upheavals (okay) \\
            African rulers' alliances resulted in rival community attacks (okay) \\
            African rulers' trade led to increased internal slavery (okay) \\
\bottomrule
\end{tabular}}}
\vspace{0.25cm}
    \caption{Automatically created nuggets for topic id 2024-35227 ``how did african rulers contribute to the triangle trade'' using documents considered by \umb to be relevant.}
\label{tab:nuggets1}
\end{table}

Our AutoNuggetizer framework represents a ``refactoring'' of this nugget evaluation methodology, brought up to date with respect to the latest advances in LLMs and the LLM-as-a-judge literature.
While there have been previous attempts to automate nugget evaluations~\citep{Lin_Demner-Fushman_HLT-EMNLP2005,Lin_Demner-Fushman_IR2006}, this effort represents the first since the advent of modern LLMs, which provide capabilities that did not previously exist.

As a concrete example, the nuggets for topic \texttt{2024-35227} ``how did african rulers contribute to the triangle trade'' are shown in Table~\ref{tab:nuggets1}.
In this case, the nuggets were automatically created from documents considered by \umb to be at least related to the topic; more details below, and see~\citet{Upadhyay:2411.08275:2024} for a complete account of the relevance assessment process.
These nuggets capture elements of what should be in a good answer, divided into ``vital'' and ``okay'' categories.
``Vital'' nuggets are those, as \citet{Voorhees_TREC2003} articulates, must be present in a good response, while ``okay'' nuggets are ``good to have'', but are not absolutely necessary.

Thus, the nugget evaluation methodology comprises two main steps:

\paragraph{\bf Nugget Creation.}
This corresponds to the first step in the nugget evaluation methodology described by \citet{Voorhees_TREC2003}, often dubbed ``nuggetization''.
Nuggets must be created based on an input set of documents under consideration; the input set is a design choice that we detail below.
Note that part of the nugget creation process is the determination of whether a nugget is ``vital'' or ``okay''.

In the original implementation of the evaluation methodology in 2003, NIST assessors manually formulated these nuggets based on documents in the pool that they assessed to be relevant.
That is, nugget creation followed relevance assessment (via pooling).
In our AutoNuggetizer framework, we have attempted to automate this process, using LLMs to generate what we call \autonuggets (see Section~\ref{section:auto-nuggets}).

It is important to note that while nuggets manifest as short natural language phrases or sentences, they are formulated at the semantic or conceptual level, and thus may or may not correspond to phrases or other lexical realizations in source documents.

\paragraph{\bf Nugget Assignment.}
This corresponds to the second step in the nugget evaluation methodology described by \citet{Voorhees_TREC2003}.
After the nuggets have been created, the list can be treated like ``an answer key''.
The assessor then reads the answer of each system to perform nugget assignment, which is a determination of whether each nugget appears in the response.
That is:\ Does a system's answer contain this particular nugget?
In our AutoNuggetizer framework, nugget assignment is performed automatically by an LLM (see Section~\ref{section:auto-assign}).

It is important to note that the nugget assignment process is performed at the semantic or conceptual level, not merely based on lexical matching (and hence requires understanding, inference, and reasoning).
In particular, a nugget can be assigned to an answer (i.e., appears in the answer) even if there is no lexical overlap between the nugget itself and the system's output.

\medskip

After the nugget assignment process, we arrive at a record of which nuggets were found in which systems' answers.
At this point, it is straightforward to compute various metrics to quantify the quality of a response and the overall score of a run.
We refer interested readers to \citet{Voorhees_TREC2003} for details on how final scores were computed previously; we adopt a different approach to quantifying answer quality in our evaluation (see Section~\ref{scoring}).

Within our general AutoNuggetizer framework, we can instantiate a number of variants.
In this work, we considered two conditions:

\begin{itemize}[leftmargin=0.5cm]

\item Automatic nugget creation with manual post-editing (\autonuggetsedits) and manual nugget assignment (\manulassign).
In this condition, we start with the set of documents that have been manually judged by NIST assessors to be relevant, which serves as the input to \autonuggets to create nuggets automatically (using an LLM, as described in Section~\ref{section:auto-nuggets}).
These nuggets are then post-edited by NIST assessors, who may add, eliminate, or combine nuggets.
Nugget creation here represents a semi-manual process where the LLM makes suggestions, but the overall process remains very much driven by human assessors.
Once these nuggets are created, the nugget assignment process is performed manually by the same NIST assessor (see Section~\ref{section:manual-assign}).
That is, the assessor reads each system's response to determine which nuggets are contained in it.

\item Automatic nugget creation (\autonuggets) and automatic nugget assignment (\autoassign).
In this condition, we start with the set of documents that have been assessed as relevant by \umb, which is a fully automated process.
Nuggets are then automatically created from this set, using \autonuggets (see Section~\ref{section:auto-nuggets}).
Given the nuggets, the nugget assignment process is also fully automatic, using \autoassign (see Section~\ref{section:auto-assign}).
Thus, this condition is fully automatic, end to end.

\end{itemize}

Details are provided below.

\begin{figure*}[p]
\begin{mdframed}[font=\small, roundcorner=10pt, linecolor=blue, linewidth=1pt, innerleftmargin=10pt, innerrightmargin=10pt, innertopmargin=10pt, innerbottommargin=10pt]
\textbf{SYSTEM:} You are NuggetizeLLM, an intelligent assistant that can update a list of atomic nuggets to best provide all the information required for the query.

\textbf{USER:} Update the list of atomic nuggets of information (1-12 words), if needed, so they best provide the information required for the query. Leverage only the initial list of nuggets (if exists) and the provided context (this is an iterative process).  Return only the final list of all nuggets in a Pythonic list format (even if no updates). Make sure there is no redundant information. Ensure the updated nugget list has at most 30 nuggets (can be less), keeping only the most vital ones. Order them in decreasing order of importance. Prefer nuggets that provide more interesting information.

Search Query: \{$query$\}

Context:

[1] \{$seg_1$\}

[2] \{$seg_2$\}

\ldots

[10] \{$seg_{10}$\}

Search Query: \{$query$\}

Initial Nugget List: \{$n_{i-1}$\}

Initial Nugget List Length: \{$len(n_{i-1})$\}

Only update the list of atomic nuggets (if needed, else return as is). Do not explain. Always answer in short nuggets (not questions). List in the form ["a", "b", ...] and a and b are strings with no mention of ".

Updated Nugget List:

\textbf{LLM:} [$n_{i+1, 1}$, \ldots]
\end{mdframed}
\caption{Prompt for the iterative LLM-based nuggetization at turn $i$.}
\label{fig:prompt.itnuggetizer}
\end{figure*}

\begin{figure*}[p]
\begin{mdframed}[font=\small, roundcorner=10pt, linecolor=blue, linewidth=1pt, innerleftmargin=10pt, innerrightmargin=10pt, innertopmargin=10pt, innerbottommargin=10pt]
\textbf{SYSTEM:} You are NuggetizeScoreLLM, an intelligent assistant that can label a list of atomic nuggets based on their importance for a given search query.

\textbf{USER:} Based on the query, label each of the \{$len(n_{f})$\} nuggets either a vital or okay based on the following criteria. Vital nuggets represent concepts that must be present in a “good” answer; on the other hand, okay nuggets contribute worthwhile information about the target but are not essential. Return the list of labels in a Pythonic list format (type: List[str]). The list should be in the same order as the input nuggets. Make sure to provide a label for each nugget.

Search Query: \{$query$\}

Nugget List: \{$n_{f}$\}

Only return the list of labels (List[str]). Do not explain.

Labels:

\textbf{LLM:} ["vital", "okay", \ldots]
\end{mdframed}
\caption{Prompt for determining the importance of nuggets. At each turn, at most 10 nuggets are passed to the LLM.}
\label{fig:prompt.nuggetscorer}
\end{figure*}


\subsection{Automatic Nugget Creation}
\label{section:auto-nuggets}

The first step in our AutoNuggetizer framework is to extract a list of atomic information units that we call nuggets from a set of input documents.
This nugget creation process, sometimes dubbed ``nuggetization'', is crucial to characterize the information that should be contained in a high-quality answer to a user query.

To perform nuggetization, we employ GPT-4o through the Azure endpoint.
The nuggetization process is run over all documents that are judged to be at least ``related'' to the query (grade $\geq$ 1).
Note that, depending on the actual evaluation condition (see above), these relevance judgments are either provided by NIST assessors or by \umb~\citep{Upadhyay_etal_arXiv2024}.
Details about relevance assessments are discussed in~\citet{Upadhyay:2411.08275:2024}

The LLM is prompted to update a list of nuggets that collectively represent the key information elements required to fully answer the query, conditioned on the provided context (documents). 
The first iteration for each query starts out with an empty pythonic list.
Our prompt design encourages the model to produce comprehensive and diverse nuggets, ensuring broad coverage of different aspects of the user's information need.
This iterative approach allows for the generation of a rich set of nuggets, capturing both explicit and implicit information requirements derived from the query.
We aim to generate nuggets that are neither too broad nor too specific.
Informed by~\citet{Voorhees_TREC2003} and previous implementations of nugget evaluations, we limit generation to at most 30 nuggets.

Once we have generated a set of nuggets for a given query, the next step is to assess the importance of each nugget. 
We again use GPT-4o; following \citet{Voorhees_TREC2003}, the LLM is asked to assign an importance label of either ``vital'' or ``okay'' with the prompt shown in Figure~\ref{fig:prompt.nuggetscorer}.

At this point, we sort the nuggets in descending order of importance and select the first 20 nuggets.
This approach usually trims a few okay nuggets and, less frequently, some vital nuggets (when there are over 20 of them).
Note that these nuggets are ordered by decreasing importance, as imposed by the prompt in Figure~\ref{fig:prompt.itnuggetizer}.
The resultant nugget set, we dub \autonuggets.

\begin{table}[t]
\setlength{\tabcolsep}{3pt}

\renewcommand{\arraystretch}{1}
\centering
\small
\resizebox{0.95\textwidth}{!}{
{\begin{tabular}{p{16cm}}
\toprule
        {\bf Automatic nugget creation (\autonuggets) using NIST qrels} \\
\midrule
            African rulers captured and sold slaves to European traders (vital) \\
            African rulers exchanged slaves for firearms and goods (vital) \\
            African rulers' involvement was crucial for the transatlantic slave trade (vital) \\
            African rulers' cooperation enabled large-scale slave trade (vital) \\
            African rulers sold war captives, criminals, and debtors (vital) \\
            African rulers benefited from the slave trade (vital) \\
            African rulers waged wars to capture more slaves (vital) \\
            African rulers' complicity was essential for the slave trade's scale (vital) \\
            African rulers' dominance over the interior facilitated the trade (vital) \\
            African rulers' involvement led to human trafficking on an industrial scale (vital) \\
            African rulers' participation was motivated by access to European goods (vital) \\
            African rulers' participation increased their wealth and power (okay) \\
            African rulers' actions had a lasting negative impact on Africa (okay) \\
            African rulers received European goods for slaves (okay) \\
            African rulers transported captives to coastal slave forts (okay) \\
            African rulers formed alliances with European traders (okay) \\
            African rulers' actions were influenced by existing African slavery practices (okay) \\
            African rulers demanded consumer articles and gold for captives (okay) \\
            African rulers encouraged European traders to come to their ports (okay) \\[1.5ex]
        \toprule
        {\bf Post-edited nuggets (\autonuggetsedits) using NIST qrels} \\
        \midrule
            African rulers sold slaves to European traders (vital) \\
            African rulers transported captives to coastal slave forts (vital) \\
            African rulers formed alliances with European traders (vital) \\
            African rulers' cooperation enabled large-scale slave trade (vital) \\
            African rulers' dominance over the interior facilitated the trade (vital) \\
            African rulers encouraged European traders to come to their ports (vital) \\
            African rulers sold war captives to European traders (okay) \\
            African rulers sold criminals to European traders (okay) \\
            African rulers sold debtors to European traders (okay) \\
            African rulers waged wars to capture more slaves (okay) \\
            African rulers conducted raids to capture more slaves (okay) \\
            African rulers received consumer goods for slaves (okay) \\
            African rulers received firearms for slaves (okay) \\
            African rulers benefited from the slave trade (okay) \\
            African rulers' participation increased their wealth (okay) \\
            African rulers' participation increased their power (okay) \\
            African rulers demanded gold for captives (okay) \\
            African rulers' actions had a lasting negative impact on Africa (okay) \\
\bottomrule
\end{tabular}}}
\vspace{0.25cm}
    \caption{Comparison of nuggets before and after human post-editing.}
\label{tab:nuggets2}
\end{table}

\subsection{Semi-Manual Nugget Creation}

In the condition that we denote as \autonuggetsedits, NIST assessors post-edit nuggets that have been proposed by \autonuggets.
Here, we start with the set of documents that have been manually judged by NIST assessors to be at least ``related'' to the query (grade $\geq 1$), which serves as the input to \autonuggets to create nuggets automatically (see Section~\ref{section:auto-nuggets}).
Note that in this case, the input set of documents has already been judged by humans to be at least related to the user's query, unlike with the \umb labels, which are generated automatically.

\begin{figure*}[p]
\begin{mdframed}[font=\small, roundcorner=10pt, linecolor=blue, linewidth=1pt, innerleftmargin=10pt, innerrightmargin=10pt, innertopmargin=10pt, innerbottommargin=10pt]
\textbf{SYSTEM:} You are NuggetizeAssignerLLM, an intelligent assistant that can label a list of atomic nuggets based on if they are captured by a given passage.

\textbf{USER:} Based on the query and passage, label each of the \{$len(n_f)$\} nuggets either as support, partial\_support, or not\_support using the following criteria. A nugget that is fully captured in the passage should be labeled as support. A nugget that is partially captured in the passage should be labeled as partial\_support. If the nugget is not captured at all, label it as not\_support. Return the list of labels in a Pythonic list format (type: List[str]). The list should be in the same order as the input nuggets. Make sure to provide a label for each nugget.

Search Query: \{$query$\}

Passage: \{$p$\}

Nugget List: \{$n_{f}$\}

Only return the list of labels (List[str]). Do not explain.

Labels:

\textbf{LLM:} ["support", "not\_support", "partial\_support", \ldots]
\end{mdframed}
\caption{Prompt for nugget assignment. At each turn, at most 10 nuggets are passed to the LLM.}
\label{fig:prompt.nuggetassigner}
\end{figure*}

The automatically generated nuggets that we provide to NIST assessors are prepared in a slightly different way:\ for each query, we create a set of 30 nuggets without any indication of importance.
This over-generation is deliberate.
These nuggets are then edited by NIST assessors, who may add, eliminate, or combine nuggets.
The NIST assessors perform this task by concurrently considering the list of relevant documents for that topic.
The process takes roughly an hour per topic, which suggests that the NIST assessors are not merely ``rubber stamping'' automatically generated nuggets, but are actually carefully deliberating over the formulation of the nuggets.

An example of this process is shown in Table~\ref{tab:nuggets2} for our running example.
On the top, we have the automatically generated nuggets, and on the bottom, we have the post-edited nuggets.

\subsection{Automatic Nugget Assignment}
\label{section:auto-assign}

The final component of our AutoNuggetizer framework, \autoassign, automatically assigns nuggets to systems' answers.
We adopt a listwise approach to nugget assignment, where the LLM is used to analyze an answer and determine if each nugget is fully supported (\texttt{support}), partially supported (\texttt{partial\_support}), or not supported (\texttt{not\_support}) by the answer.
Again, we employ GPT-4o through the Azure endpoint, with the prompt shown in Figure~\ref{fig:prompt.nuggetassigner}.
We iteratively prompt the model with at most 10 nuggets to evaluate assignment at each step.

\begin{table}[t]
\setlength{\tabcolsep}{3pt}
\small
\renewcommand{\arraystretch}{1}
\centering
\resizebox{0.95\textwidth}{!}{
{\begin{tabular}{p{13cm}r}
\toprule
        \textbf{Nuggets} & \textbf{Assignment} \\
        \midrule
        {\bf Automatic nugget creation (\autonuggets) using \umb qrels, \autoassign } \\ 
        African rulers captured and sold slaves to Europeans (vital) & Support \\
        African rulers waged wars to capture more slaves (vital) & Not Support \\
        African rulers exchanged slaves for firearms (vital) & Partial Support \\
        African rulers' involvement was crucial for the trade's scale (vital) & Support \\
        African rulers' involvement increased due to European demand (vital) & Partial Support \\
        African rulers captured slaves from raids and wars (vital) & Partial Support \\
        African rulers' control and supply of captives was significant (vital) & Support \\
        African rulers transported captives to coastal slave forts (vital) & Support \\
        African rulers traded slaves for textiles and ironware (vital) & Not Support \\
        African rulers grew wealthy from the slave trade (okay) & Support \\
        African rulers' alliances with Europeans were fundamental (okay) & Support \\
        African rulers' trade caused increased tension and violence (okay) & Partial Support \\
        African rulers' actions led to internal African upheavals (okay) & Partial Support \\
        African rulers' alliances resulted in rival community attacks (okay) & Partial Support \\
        African rulers' trade led to increased internal slavery (okay) & Partial Support \\
        \midrule
        {\bf Post-edited nuggets (\autonuggetsedits) using NIST qrels, \manulassign} \\
        African rulers sold slaves to European traders (vital) & Support \\
        African rulers sold war captives to European traders (okay) & Not Support \\
        African rulers sold criminals to European traders (okay) & Not Support \\
        African rulers sold debtors to European traders (okay) & Not Support \\
        African rulers transported captives to coastal slave forts (vital) & Not Support \\
        African rulers waged wars to capture more slaves (okay) & Not Support \\
        African rulers conducted raids to capture more slaves (okay) & Not Support \\
        African rulers formed alliances with European traders (vital) & Not Support \\
        African rulers' cooperation enabled large-scale slave trade (vital) & Not Support \\
        African rulers' dominance over the interior facilitated the trade (vital) & Not Support \\
        African rulers encouraged European traders to come to their ports (vital) & Not Support \\
        African rulers received consumer goods for slaves (okay) & Support \\
        African rulers received firearms for slaves (okay) & Support \\
        African rulers benefited from the slave trade (okay) & Support \\
        African rulers' participation increased their wealth (okay) & Not Support \\
        African rulers' participation increased their power (okay) & Not Support \\
        African rulers demanded gold for captives (okay) & Support \\
        African rulers' actions had a lasting negative impact on Africa (okay) & Not Support \\
        \bottomrule
    \end{tabular}}}
    \vspace{0.25cm}
    \caption{Comparison of nugget assignment and nugget creation approaches.}
    \label{tab:assignment}
\end{table}

\subsection{Manual Nugget Assignment}
\label{section:manual-assign}

In the manual nugget assignment process, we leverage the expertise of the original NIST annotator involved in creating the list of nuggets. 
The annotator examines each answer text, assigning each nugget one of three labels based on the extent of its support: \texttt{support}, \texttt{partial\_support}, and \texttt{not\_support} (same as with automatic assignment). 
It is worth clarifying that annotators in this step are not shown any LLM output, so this is a fully manual process.

By involving the same annotator responsible for nugget creation, we ensure continuity and consistency in nugget interpretation.
This continuity helps maintain reliability across the nugget assignment process, as the annotator applies the original context and intent behind each nugget to the assessment of answer coverage.

An example is shown in Table~\ref{tab:assignment}:\ on the top, we show fully automatic assignment (\autoassign) of automatically created nuggets (\autonuggets) using \umb qrels; on the bottom, we show manual assignment (\manulassign) of post-edited nuggets (\autonuggetsedits) using NIST qrels.
The answer being evaluated is the one shown in Table~\ref{tab:sample-sample}.

\subsection{Scoring}
\label{scoring}

At this point in the evaluation, we have already completed nugget creation and nugget assignment.
For each query, we have a list of nuggets, and for each system's response, we have a record of which nuggets it contains, in terms of a three-way judgment:\ \texttt{support}, \texttt{partial\_support}, and \texttt{not\_support} (either manual or automatic nugget assignment).

The final step is to compute the score for a system's response to a query $q$.
The score of a run is simply the mean of the score across all queries.
We compute the following scores per query:

\paragraph{\textbf{All} ($A$)}
This score is the average of the scores for all nuggets in an answer.
Given:
\begin{align}
    s_i &= \begin{cases}
        1 & \text{if assignment = \texttt{support}} \\
        0.5 & \text{if assignment = \texttt{partial\_support}} \\
        0 & \text{otherwise}
    \end{cases}
\end{align}
The ``All'' score is then calculated as:
$$A = \frac{\sum_{i} s_i}{N_\text{nuggets}}$$

\paragraph{\textbf{All Strict} ($A_{\textrm{strict}}$)}
This variant is calculated the same as the ``All'' score above, but with stricter nugget matching:
\begin{align}
    ss_i &= \begin{cases}
        1 & \text{if assignment = \texttt{support}} \\
        0 & \text{otherwise}
    \end{cases}
\end{align}
$$A_{\textrm{strict}} = \frac{\sum_i ss_i}{N_\text{nuggets}}$$

\paragraph{\textbf{Vital} ($V$)}
This score is the average of the scores for only the vital nuggets in the answer, where $\mathrm{n^v}$ represents the subset of the vital nuggets.
We define $s^v_i$ in the same exact way as $s_i$ above, but only over the vital nuggets:

$$V = \frac{\sum_i s^v_i}{|\mathrm{n^v}|}$$

\paragraph{\textbf{Vital Strict} ($V_{\textrm{strict}}$)} This score is used as the primary metric in the evaluation.
It is calculated the same as $V$, but with stricter nugget matching.
We define $ss^v_i$ in the same exact way as $ss_i$ above, but only over the vital nuggets:

$$V_{\textrm{strict}} = \frac{\sum_i ss_i^v}{|\mathbf{n^v}|}$$

\paragraph{\textbf{Weighted Score} ($W$)}
Here we assign a weight of 1 to vital nuggets and 0.5 to okay nuggets.
We define $\mathrm{n^v}$ to be the vital nuggets and $\mathrm{n^o}$ to be the okay nuggets.
The score $s^v_i$ is defined the same as above; the score $s^o_i$ is defined similarly, but over okay nuggets.
Then:

$$W = \frac{\sum s^v_i + 0.5 \cdot \sum s^o_i}{|\mathbf{n^v}| + 0.5 \cdot |\mathbf{n^o}|}$$

\paragraph{\textbf{Weighted Score Strict} ($W_{\textrm{strict}}$)}
This variant is calculated the same way as the Weighted Score, but using the stricter nugget matching, i.e., $ss^v_i$ and its counterpart $ss^o_i$.

$$W_{\textrm{strict}} = \frac{\sum ss^v_i + 0.5 \cdot \sum ss^o_i}{|\mathbf{n^v}| + 0.5 \cdot |\mathbf{n^o}|}$$

\begin{table*}[p]
\centering
\scalebox{0.72}{
\begin{tabular}{lrccccccr}
\toprule
Run ID & Task & $V_{\textrm{strict}}$ &  $V$ & $W_{\textrm{strict}}$ & $W$ & $A_{\textrm{strict}}$ & $A$ & $L$ \\
\midrule
ldisnu.ldilab\_gpt\_4o                   & AG & 0.6666 & 0.6758 & 0.5947 & 0.6113 & 0.5581 & 0.5784 & 290.86 \\
h2oloo.listgalore\_gpt4o\_ragnarokv4\_top20 & RAG & 0.6411 & 0.6589 & 0.5785 & 0.6044 & 0.5483 & 0.5764 & 287.71 \\
neu.neuragfix                            & RAG & 0.6372 & 0.6534 & 0.5896 & 0.6138 & 0.5626 & 0.5906 & 327.62 \\
neu.neurag                               & RAG & 0.6133 & 0.6243 & 0.5745 & 0.5931 & 0.5511 & 0.5734 & 327.62 \\
ldisnu.dilab\_repllama\_listt5\_pass3\_gpt4o & RAG & 0.6121 & 0.6217 & 0.5681 & 0.5764 & 0.5425 & 0.5497 & 291.90 \\
coordinators.baseline\_frag\_rag24.test\_gpt-4o\_top20 & RAG & 0.6074 & 0.6284 & 0.5541 & 0.5787 & 0.5242 & 0.5493 & 306.43 \\
softbank-meisei.agtask-bm25-colbert\_faiss-gpt4o-llama70b & AG & 0.5497 & 0.5685 & 0.5124 & 0.5297 & 0.4901 & 0.5060 & 213.81 \\
KML.gpt\_mini                            & AG & 0.5476 & 0.5605 & 0.5281 & 0.5456 & 0.5148 & 0.5345 & 373.62 \\
CIR.cir\_gpt-4o-mini\_no\_reranking\_50\_0.5\_100\_301\_p1 & AG & 0.5298 & 0.5509 & 0.4990 & 0.5248 & 0.4794 & 0.5073 & 151.90 \\
coordinators.baseline\_frag\_rag24.test\_command-r-plus\_top20 & RAG & 0.5261 & 0.5366 & 0.4963 & 0.5141 & 0.4768 & 0.4975 & 234.05 \\
ncsu-las.LAS-splade-mxbai-rrf-mmr8       & RAG & 0.5254 & 0.5452 & 0.4825 & 0.5014 & 0.4638 & 0.4817 & 299.33 \\
coordinators.baseline\_rag24.test\_l31\_70b\_instruct\_top20 & AG & 0.5218 & 0.5326 & 0.4558 & 0.4720 & 0.4274 & 0.4462 & 194.10 \\
webis.webis-rag-run1-taskrag             & RAG & 0.5131 & 0.5386 & 0.4769 & 0.5002 & 0.4591 & 0.4806 & 177.62 \\
CIR.cir\_gpt-4o-mini\_Cosine\_50\_0.5\_100\_301\_p1 & AG & 0.5104 & 0.5176 & 0.4831 & 0.4951 & 0.4640 & 0.4784 & 161.57 \\
KML.cohere+post\_processing              & AG & 0.5055 & 0.5154 & 0.4609 & 0.4707 & 0.4370 & 0.4466 & 342.57 \\
WaterlooClarke.UWCrag                    & RAG & 0.4856 & 0.4894 & 0.4407 & 0.4493 & 0.4194 & 0.4300 & 170.43 \\
buw.buw                                  & RAG & 0.4806 & 0.4899 & 0.4174 & 0.4258 & 0.3880 & 0.3954 & 193.86 \\
ncsu-las.LAS-splade-mxbai-mmr8-RAG       & RAG & 0.4637 & 0.4807 & 0.4500 & 0.4670 & 0.4316 & 0.4484 & 296.05 \\
softbank-meisei.rag\_bm25-colbert\_faiss-gpt4o-llama70b & RAG & 0.4610 & 0.4814 & 0.4263 & 0.4450 & 0.4046 & 0.4216 & 179.10 \\
Ruc01.ruc001                             & RAG & 0.4479 & 0.4640 & 0.3916 & 0.4054 & 0.3620 & 0.3744 & 236.48 \\
h2oloo.listgalore\_l31-70b\_ragnarokv4\_top20 & RAG & 0.4477 & 0.4594 & 0.3960 & 0.4137 & 0.3701 & 0.3910 & 208.24 \\
citi.BEST\_cot\_gpt3.5                   & RAG & 0.4448 & 0.4529 & 0.3961 & 0.4074 & 0.3690 & 0.3817 & 198.86 \\
webis.webis-rag-run0-taskrag             & RAG & 0.4407 & 0.4584 & 0.4148 & 0.4342 & 0.3989 & 0.4185 & 162.95 \\
InfoLab.UDInfolab.RAG.Query              & RAG & 0.4383 & 0.4489 & 0.4181 & 0.4319 & 0.4014 & 0.4165 & 149.33 \\
InfoLab.UDInfolab.RAG.AnsAI              & RAG & 0.4322 & 0.4455 & 0.4099 & 0.4249 & 0.3905 & 0.4057 & 158.05 \\
IITD-IRL.ag\_rag\_gpt35\_expansion\_rrf\_20 & AG & 0.4282 & 0.4387 & 0.4004 & 0.4091 & 0.3817 & 0.3890 & 97.19 \\
ielab.ielab-b70bf-70bqfs-ad\_hoc         & RAG & 0.4240 & 0.4331 & 0.3740 & 0.3888 & 0.3500 & 0.3673 & 180.29 \\
TREMA-UNH.Ranked\_Iterative\_Fact\_Extraction\ldots & AG & 0.4014 & 0.4176 & 0.3678 & 0.3885 & 0.3522 & 0.3753 & 373.05 \\
citi.SECOND\_cot\_gpt3.5                 & RAG & 0.4008 & 0.4123 & 0.3628 & 0.3776 & 0.3395 & 0.3556 & 171.10 \\
TREMA-UNH.Enhanced\_Iterative\_Fact\_Refinement\ldots & AG & 0.3971 & 0.4143 & 0.3773 & 0.3953 & 0.3662 & 0.3853 & 379.95 \\
IITD-IRL.zeph\_test\_rag\_rrf\_expand\_query & RAG & 0.3826 & 0.3958 & 0.3422 & 0.3547 & 0.3227 & 0.3344 & 84.48 \\
uis-iai.baseline\_top\_5                 & AG & 0.3785 & 0.3909 & 0.3394 & 0.3500 & 0.3194 & 0.3285 & 82.43 \\
WaterlooClarke.UWCgarag                  & RAG & 0.3733 & 0.3945 & 0.3535 & 0.3723 & 0.3435 & 0.3605 & 223.48 \\
buw.oneshot\_post\_sentenced             & RAG & 0.3563 & 0.3682 & 0.3256 & 0.3392 & 0.3087 & 0.3221 & 214.19 \\
IIIA-UNIPD.iiia\_standard\_p1\_straight\_ag & AG & 0.3491 & 0.3591 & 0.3289 & 0.3403 & 0.3146 & 0.3264 & 133.48 \\
uis-iai.ginger\_top\_5                   & AG & 0.3256 & 0.3390 & 0.2953 & 0.3169 & 0.2803 & 0.3050 & 102.71 \\
ielab.ielab-b70bf-70bqp-70bafs           & RAG & 0.3067 & 0.3155 & 0.2681 & 0.2780 & 0.2496 & 0.2598 & 84.05 \\
uog-tht.FT-llama3                        & RAG & 0.3017 & 0.3280 & 0.2754 & 0.2991 & 0.2642 & 0.2861 & 79.43 \\
uog-tht.ICL-mistral                      & RAG & 0.2899 & 0.3128 & 0.2698 & 0.2887 & 0.2608 & 0.2769 & 79.48 \\
IIIA-UNIPD.iiia\_dedup\_p1\_straight\_ag & AG & 0.2799 & 0.2924 & 0.2604 & 0.2719 & 0.2525 & 0.2629 & 91.24 \\
SGU.qrant\_bge\_gemini                   & RAG & 0.2546 & 0.2599 & 0.2071 & 0.2143 & 0.1862 & 0.1948 & 128.10 \\
ii\_research.iiresearch-bm25-top10-llama3-8b-instruct & RAG & 0.1999 & 0.2043 & 0.1798 & 0.1850 & 0.1672 & 0.1724 & 178.76 \\
IRIT.ISIR-IRIT-zephyr\_query\_gen        & RAG & 0.1546 & 0.1868 & 0.1433 & 0.1725 & 0.1358 & 0.1617 & 151.62 \\
IRIT.ISIR-IRIT-zephyr\_p2                & RAG & 0.1469 & 0.1726 & 0.1378 & 0.1595 & 0.1290 & 0.1474 & 148.10 \\
webis.webis-manual                       & RAG & 0.0640 & 0.0807 & 0.0752 & 0.0918 & 0.0787 & 0.0942 & 29.24 \\
\midrule
\textbf{Min}                             &  & 0.0079 & 0.0079 & 0.0117 & 0.0127 & 0.0121 & 0.0134 & 9.95 \\
\textbf{Median}                          &  & 0.4126 & 0.4234 & 0.3962 & 0.4119 & 0.3759 & 0.3978 & 179.71 \\
\textbf{Max}                             &  & 0.8148 & 0.8258 & 0.7480 & 0.7650 & 0.7185 & 0.7385 & 390.00 \\
\bottomrule
\end{tabular}}
\vspace{0.25cm}
\caption{Scores for AG and RAG runs under the \autonuggetsedits { / }\manulassign condition, over 21 topics that have been manually evaluated by NIST.}
\label{tab:edit_nugget_nist_assignment_all}
\end{table*}
\begin{table*}[p]
\centering
\scalebox{0.72}{
\begin{tabular}{lrccccccr}
\toprule
Run ID & Task & $V_{\textrm{strict}}$ &  $V$ & $W_{\textrm{strict}}$ & $W$ & $A_{\textrm{strict}}$ & $A$ & $L$ \\
\midrule
h2oloo.listgalore\_gpt4o\_ragnarokv4\_top20 & RAG & 0.4772 & 0.5965 & 0.4665 & 0.5868 & 0.4623 & 0.5818 & 287.71 \\
neu.neuragfix                            & RAG & 0.4456 & 0.5616 & 0.4242 & 0.5405 & 0.4144 & 0.5300 & 327.62 \\
neu.neurag                               & RAG & 0.4345 & 0.5502 & 0.4176 & 0.5320 & 0.4097 & 0.5230 & 327.62 \\
coordinators.baseline\_frag\_rag24.test\_gpt-4o\_top20 & RAG & 0.4340 & 0.5614 & 0.4245 & 0.5492 & 0.4221 & 0.5440 & 306.43 \\
CIR.cir\_gpt-4o-mini\_no\_reranking\_50\_0.5\_100\_301\_p1 & AG & 0.4254 & 0.5531 & 0.4044 & 0.5292 & 0.3957 & 0.5178 & 151.90 \\
softbank-meisei.agtask-bm25-colbert\_faiss-gpt4o-llama70b & AG & 0.4245 & 0.5433 & 0.4031 & 0.5245 & 0.3930 & 0.5145 & 213.81 \\
ldisnu.dilab\_repllama\_listt5\_pass3\_gpt4o & RAG & 0.4171 & 0.5330 & 0.3965 & 0.5193 & 0.3840 & 0.5105 & 291.90 \\
CIR.cir\_gpt-4o-mini\_Cosine\_50\_0.5\_100\_301\_p1 & AG & 0.4164 & 0.5163 & 0.3970 & 0.5012 & 0.3863 & 0.4924 & 161.57 \\
ldisnu.ldilab\_gpt\_4o                   & AG & 0.4153 & 0.5512 & 0.3961 & 0.5292 & 0.3873 & 0.5182 & 290.86 \\
webis.webis-rag-run1-taskrag             & RAG & 0.4050 & 0.5249 & 0.3879 & 0.5085 & 0.3792 & 0.4999 & 177.62 \\
KML.gpt\_mini                            & AG & 0.4000 & 0.5516 & 0.3681 & 0.5271 & 0.3533 & 0.5161 & 373.62 \\
webis.webis-rag-run0-taskrag             & RAG & 0.3906 & 0.4850 & 0.3800 & 0.4783 & 0.3751 & 0.4756 & 162.95 \\
ncsu-las.LAS-splade-mxbai-mmr8-RAG       & RAG & 0.3886 & 0.4877 & 0.3676 & 0.4711 & 0.3582 & 0.4636 & 296.05 \\
KML.cohere+post\_processing              & AG & 0.3818 & 0.5033 & 0.3541 & 0.4858 & 0.3400 & 0.4773 & 342.57 \\
InfoLab.UDInfolab.RAG.AnsAI              & RAG & 0.3768 & 0.4951 & 0.3497 & 0.4677 & 0.3357 & 0.4538 & 158.05 \\
ncsu-las.LAS-splade-mxbai-rrf-mmr8       & RAG & 0.3729 & 0.4974 & 0.3502 & 0.4760 & 0.3411 & 0.4669 & 299.33 \\
coordinators.baseline\_frag\_rag24.test\_command-r-plus\_top20 & RAG & 0.3696 & 0.4849 & 0.3502 & 0.4702 & 0.3407 & 0.4628 & 234.05 \\
h2oloo.listgalore\_l31-70b\_ragnarokv4\_top20 & RAG & 0.3630 & 0.4718 & 0.3325 & 0.4498 & 0.3144 & 0.4368 & 208.24 \\
WaterlooClarke.UWCgarag                  & RAG & 0.3452 & 0.4894 & 0.3186 & 0.4610 & 0.3023 & 0.4447 & 223.48 \\
softbank-meisei.rag\_bm25-colbert\_faiss-gpt4o-llama70b & RAG & 0.3390 & 0.4405 & 0.3009 & 0.4123 & 0.2804 & 0.3983 & 179.10 \\
WaterlooClarke.UWCrag                    & RAG & 0.3325 & 0.4319 & 0.3070 & 0.4145 & 0.2936 & 0.4047 & 170.43 \\
InfoLab.UDInfolab.RAG.Query              & RAG & 0.3255 & 0.4467 & 0.3046 & 0.4348 & 0.2951 & 0.4292 & 149.33 \\
coordinators.baseline\_rag24.test\_l31\_70b\_instruct\_top20 & AG & 0.3213 & 0.4448 & 0.3136 & 0.4412 & 0.3088 & 0.4393 & 194.10 \\
citi.SECOND\_cot\_gpt3.5                 & RAG & 0.2999 & 0.4126 & 0.2763 & 0.3852 & 0.2652 & 0.3721 & 171.10 \\
citi.BEST\_cot\_gpt3.5                   & RAG & 0.2971 & 0.4024 & 0.2813 & 0.3940 & 0.2725 & 0.3888 & 198.86 \\
IITD-IRL.ag\_rag\_gpt35\_expansion\_rrf\_20 & AG & 0.2920 & 0.4044 & 0.2807 & 0.3909 & 0.2760 & 0.3841 & 97.19 \\
IIIA-UNIPD.iiia\_standard\_p1\_straight\_ag & AG & 0.2854 & 0.3996 & 0.2658 & 0.3856 & 0.2542 & 0.3768 & 133.48 \\
Ruc01.ruc001                             & RAG & 0.2840 & 0.3836 & 0.2819 & 0.3845 & 0.2813 & 0.3855 & 236.48 \\
ielab.ielab-b70bf-70bqfs-ad\_hoc         & RAG & 0.2781 & 0.4021 & 0.2525 & 0.3829 & 0.2380 & 0.3715 & 180.29 \\
buw.buw                                  & RAG & 0.2757 & 0.3890 & 0.2633 & 0.3737 & 0.2561 & 0.3652 & 193.86 \\
IITD-IRL.zeph\_test\_rag\_rrf\_expand\_query & RAG & 0.2673 & 0.3610 & 0.2527 & 0.3444 & 0.2449 & 0.3347 & 84.48 \\
TREMA-UNH.Ranked\_Iterative\_Fact\_Extraction\_and\ldots & AG & 0.2662 & 0.3849 & 0.2554 & 0.3760 & 0.2500 & 0.3719 & 373.05 \\
buw.oneshot\_post\_sentenced             & RAG & 0.2575 & 0.3817 & 0.2401 & 0.3650 & 0.2311 & 0.3562 & 214.19 \\
uis-iai.baseline\_top\_5                 & AG & 0.2496 & 0.3656 & 0.2294 & 0.3467 & 0.2182 & 0.3362 & 82.43 \\
TREMA-UNH.Enhanced\_Iterative\_Fact\_Refinement\ldots & AG & 0.2475 & 0.3904 & 0.2355 & 0.3778 & 0.2285 & 0.3707 & 379.95 \\
ielab.ielab-b70bf-70bqp-70bafs           & RAG & 0.2244 & 0.3281 & 0.2219 & 0.3210 & 0.2192 & 0.3154 & 84.05 \\
uis-iai.ginger\_top\_5                   & AG & 0.2109 & 0.3321 & 0.2065 & 0.3278 & 0.2052 & 0.3264 & 102.71 \\
uog-tht.ICL-mistral                      & RAG & 0.2012 & 0.3268 & 0.1828 & 0.3091 & 0.1731 & 0.2999 & 79.48 \\
uog-tht.FT-llama3                        & RAG & 0.2012 & 0.3206 & 0.1813 & 0.3041 & 0.1707 & 0.2956 & 79.43 \\
IIIA-UNIPD.iiia\_dedup\_p1\_straight\_ag & AG & 0.2004 & 0.3045 & 0.1957 & 0.2975 & 0.1919 & 0.2923 & 91.24 \\
ii\_research.iiresearch-bm25-top10-llama3-8b-instruct & RAG & 0.1378 & 0.2023 & 0.1258 & 0.1918 & 0.1196 & 0.1870 & 178.76 \\
SGU.qrant\_bge\_gemini                   & RAG & 0.1343 & 0.1975 & 0.1277 & 0.1896 & 0.1263 & 0.1873 & 128.10 \\
IRIT.ISIR-IRIT-zephyr\_query\_gen        & RAG & 0.1310 & 0.2294 & 0.1295 & 0.2243 & 0.1274 & 0.2212 & 151.62 \\
IRIT.ISIR-IRIT-zephyr\_p2                & RAG & 0.1221 & 0.2241 & 0.1233 & 0.2174 & 0.1226 & 0.2125 & 148.10 \\
webis.webis-manual                       & RAG & 0.0687 & 0.0968 & 0.0750 & 0.0995 & 0.0797 & 0.1025 & 29.24 \\
\midrule
\textbf{Min}                             &  & 0.0073 & 0.0150 & 0.0122 & 0.0190 & 0.0143 & 0.0210 & 9.95 \\
\textbf{Median}                          &  & 0.3121 & 0.4388 & 0.2967 & 0.4271 & 0.2837 & 0.4183 & 179.71 \\
\textbf{Max}                             &  & 0.5842 & 0.6949 & 0.5550 & 0.6690 & 0.5465 & 0.6607 & 390.00 \\
\bottomrule
\end{tabular}}
\vspace{0.25cm}
\caption{Scores for AG and RAG runs under the \autonuggets { / }\autoassign condition, over 21 topics that have been manually evaluated by NIST.}
\label{tab:auto_nugget_auto_assignment_all}
\end{table*}
 \begin{figure}[t]
        \centering
        \includegraphics[width=0.93\textwidth]{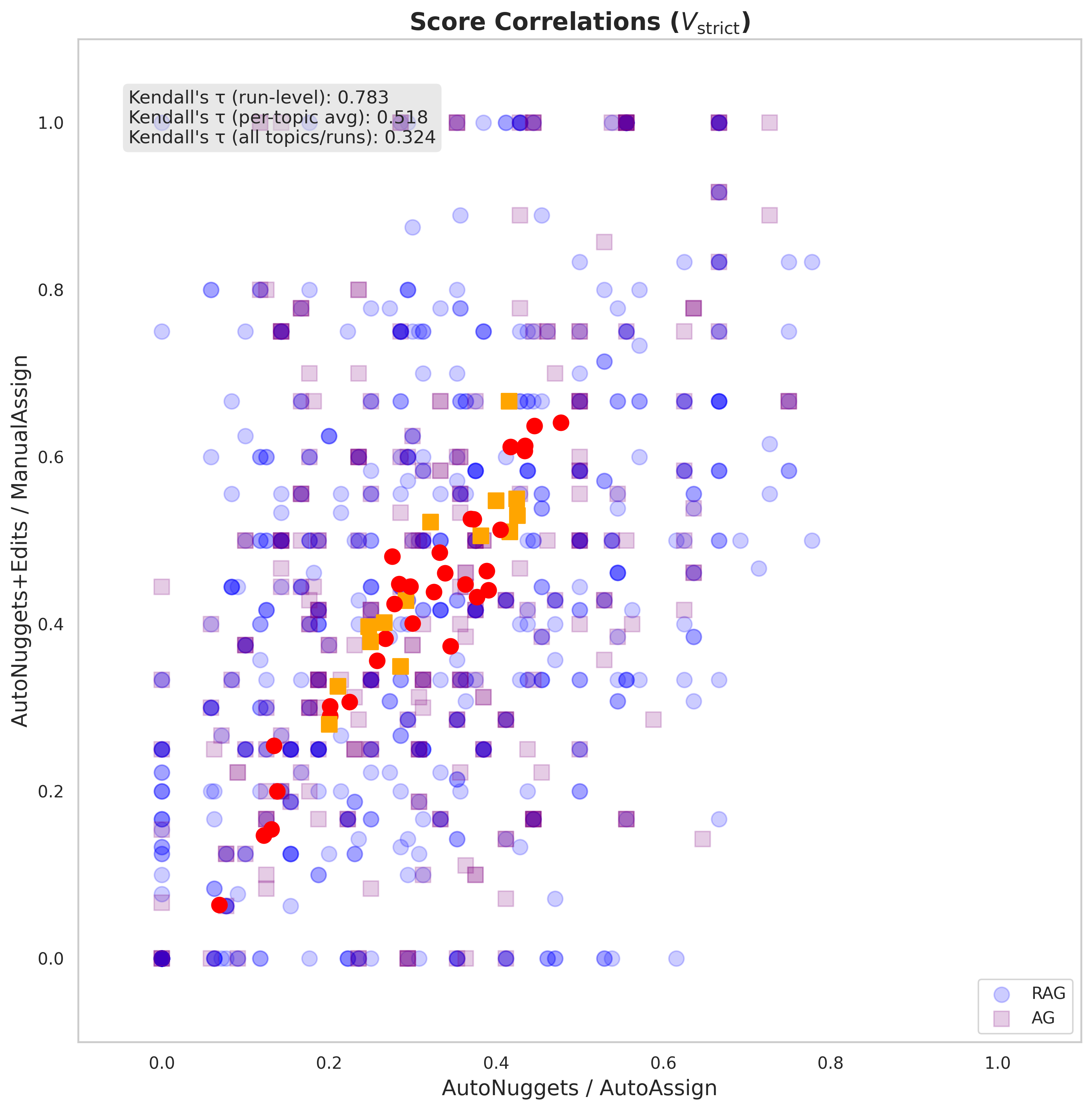}
        
    \caption{Scatter plot showing correlations between manual vs.\ automatic $V_{\textrm{strict}}$ scores for AG and RAG runs, over 21 topics that have been manually evaluated by NIST. The $x$ axis shows automatic scores from the \autonuggets { } / \autoassign condition and the $y$ axis shows manual scores from the \autonuggetsedits { }/ \manulassign condition.   
    Red circles (RAG runs) and orange squares (AG runs) represent run-level scores from Tables~\ref{tab:edit_nugget_nist_assignment_all} and~\ref{tab:auto_nugget_auto_assignment_all}. Blue circles (RAG runs) and purple squares (AG runs) show all topic/run combinations. The top-left box reports Kendall's $\tau$ correlations at the run level (red circles/orange squares), over all topic/run combinations (blue circles/purple squares), and the variant where we compute Kendall's $\tau$ per topic, and then average over the per-topic values.}
    \label{fig:scatter_plots_1}
\end{figure}
 \begin{figure}[t]
        \centering
        \includegraphics[width=0.93\textwidth]{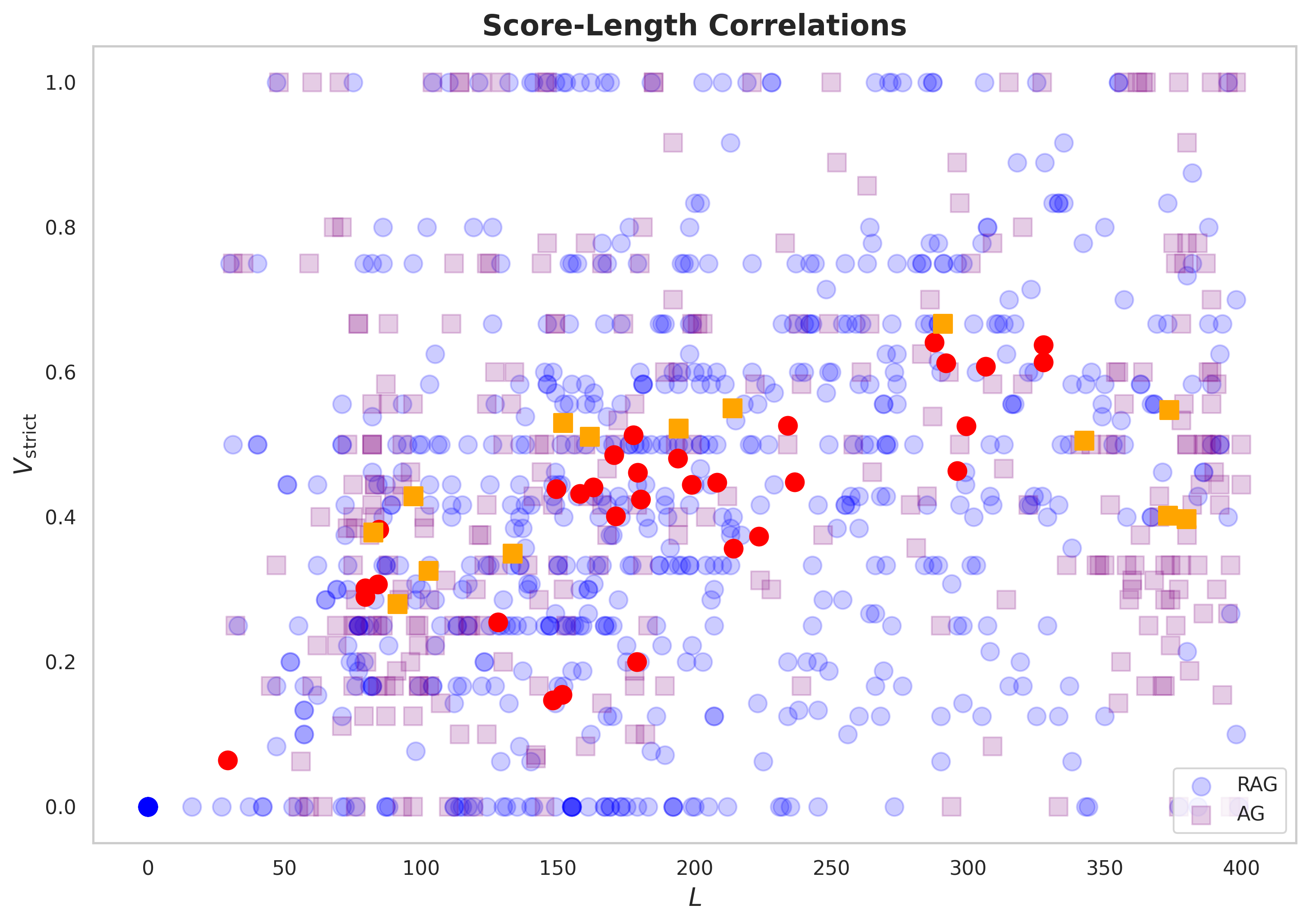}
        
    \caption{Scatter plot of $L$ and $V_{\textrm{strict}}$ from Table~\ref{tab:edit_nugget_nist_assignment_all} over 21 topics that have been manually evaluated by NIST. The red/orange dots show run-level scores and the blue/purple dots show all topic/run combinations. Circles indicate RAG submissions and squares, AG.}
    \label{fig:scatter_plots_5}
\end{figure}
\begin{table*}[p]
\centering
\scalebox{0.72}{
\begin{tabular}{lrccccccr}
\toprule
Run ID & Task & $V_{\textrm{strict}}$ &  $V$ & $W_{\textrm{strict}}$ & $W$ & $A_{\textrm{strict}}$ & $A$ & $L$ \\
\midrule
coordinators.all\_nuggets                & RAG & 0.9692 & 0.9790 & 0.9492 & 0.9640 & 0.9386 & 0.9561 & 127.54 \\
h2oloo.listgalore\_gpt4o\_ragnarokv4\_top20 & RAG & 0.4546 & 0.5840 & 0.4318 & 0.5598 & 0.4202 & 0.5472 & 287.19 \\
neu.neuragfix                            & RAG & 0.4422 & 0.5639 & 0.4162 & 0.5384 & 0.4026 & 0.5249 & 336.31 \\
neu.neurag                               & RAG & 0.4416 & 0.5647 & 0.4160 & 0.5399 & 0.4026 & 0.5266 & 336.31 \\
uis-iai.ginger-fluency\_top\_20          & AG & 0.4267 & 0.5766 & 0.4151 & 0.5627 & 0.4091 & 0.5554 & 377.55 \\
ncsu-las.LAS-T5-mxbai-mmr8-RAG           & RAG & 0.4233 & 0.5528 & 0.4027 & 0.5303 & 0.3914 & 0.5181 & 293.26 \\
softbank-meisei.agtask-bm25-colbert\_faiss-gpt4o-llama70b & AG & 0.4196 & 0.5464 & 0.3996 & 0.5236 & 0.3901 & 0.5125 & 222.42 \\
ldisnu.ldilab\_gpt\_4o                   & AG & 0.4186 & 0.5366 & 0.3965 & 0.5143 & 0.3853 & 0.5030 & 295.98 \\
coordinators.baseline\_rag24.test\_gpt-4o\_top20 & AG & 0.4173 & 0.5389 & 0.3948 & 0.5156 & 0.3830 & 0.5035 & 300.93 \\
ldisnu.dilab\_repllama\_listt5\_pass3\_gpt4o & RAG & 0.4147 & 0.5325 & 0.3938 & 0.5100 & 0.3818 & 0.4972 & 297.27 \\
coordinators.baseline\_frag\_rag24.test\_gpt-4o\_top20 & RAG & 0.4141 & 0.5348 & 0.3924 & 0.5117 & 0.3810 & 0.4997 & 300.93 \\
ncsu-las.LAS\_splad\_mxbai-rrf-occams\_50\_RAG & RAG & 0.4136 & 0.5414 & 0.3952 & 0.5213 & 0.3849 & 0.5103 & 364.57 \\
ldisnu.dilab\_repllama\_listt5\_pass4\_gpt4o & RAG & 0.4081 & 0.5215 & 0.3872 & 0.4998 & 0.3751 & 0.4873 & 294.35 \\
webis.webis-rag-run1-taskrag             & RAG & 0.4079 & 0.5437 & 0.3823 & 0.5161 & 0.3690 & 0.5016 & 183.72 \\
ldisnu.dilab\_repllama\_listt5\_pass2\_gpt4o & RAG & 0.4061 & 0.5265 & 0.3885 & 0.5059 & 0.3785 & 0.4944 & 296.85 \\
ncsu-las.LAS-splade-mxbai-rrf-mmr8-doc   & RAG & 0.4056 & 0.5263 & 0.3864 & 0.5045 & 0.3760 & 0.4928 & 288.55 \\
ldisnu.dilab\_repllama\_listt5\_pass1\_gpt4o & RAG & 0.4055 & 0.5203 & 0.3840 & 0.4980 & 0.3724 & 0.4860 & 290.04 \\
KML.gpt\_mini                            & AG & 0.4010 & 0.5478 & 0.3839 & 0.5297 & 0.3750 & 0.5202 & 373.86 \\
webis.webis-ag-run1-taskrag              & AG & 0.3991 & 0.5392 & 0.3765 & 0.5150 & 0.3647 & 0.5026 & 180.59 \\
CIR.cir\_gpt-4o-mini\_Cosine\_50\_0.5\_100\_301\_p3 & AG & 0.3982 & 0.5434 & 0.3796 & 0.5218 & 0.3689 & 0.5097 & 177.67 \\
ncsu-las.LAS-splade-mxbai-mmr8-RAG       & RAG & 0.3978 & 0.5184 & 0.3791 & 0.5002 & 0.3691 & 0.4906 & 301.26 \\
h2oloo.listgalore\_gpt4o\_ragnarokv4nocite\_top20 & RAG & 0.3964 & 0.5509 & 0.3693 & 0.5213 & 0.3545 & 0.5056 & 327.76 \\
CIR.cir\_gpt-4o-mini\_Cosine\_50\_0.75\_100\_301\_p1 & AG & 0.3903 & 0.5234 & 0.3695 & 0.5035 & 0.3583 & 0.4931 & 160.05 \\
CIR.cir\_gpt-4o-mini\_Cosine\_50\_1.0\_100\_301\_p1 & AG & 0.3880 & 0.5222 & 0.3693 & 0.5009 & 0.3594 & 0.4897 & 160.70 \\
CIR.cir\_gpt-4o-mini\_Cosine\_50\_0.5\_100\_301\_p2 & AG & 0.3875 & 0.5285 & 0.3659 & 0.5055 & 0.3537 & 0.4929 & 170.65 \\
ncsu-las.LAS-splade-mxbai-rrf-mmr8       & RAG & 0.3871 & 0.5208 & 0.3709 & 0.5041 & 0.3621 & 0.4947 & 295.69 \\
CIR.cir\_gpt-4o-mini\_no\_reranking\_50\_0.5\_100\_301\_p1 & AG & 0.3860 & 0.5210 & 0.3705 & 0.5031 & 0.3627 & 0.4943 & 159.79 \\
CIR.cir\_gpt-4o-mini\_Cosine\_50\_0.5\_100\_301\_p1 & AG & 0.3824 & 0.5269 & 0.3617 & 0.5030 & 0.3502 & 0.4901 & 157.48 \\
CIR.cir\_gpt-4o-mini\_Cosine\_50\_0.25\_100\_301\_p1 & AG & 0.3822 & 0.5226 & 0.3576 & 0.4952 & 0.3435 & 0.4801 & 158.45 \\
webis.webis-rag-run0-taskrag             & RAG & 0.3800 & 0.5102 & 0.3572 & 0.4833 & 0.3449 & 0.4693 & 165.40 \\
KML.gpt\_mini\_double\_prompt            & AG & 0.3777 & 0.5357 & 0.3545 & 0.5097 & 0.3422 & 0.4961 & 372.71 \\
uis-iai.ginger-fluency\_top\_10          & AG & 0.3690 & 0.5088 & 0.3591 & 0.4959 & 0.3539 & 0.4895 & 256.72 \\
WaterlooClarke.UWCgarag                  & RAG & 0.3664 & 0.5150 & 0.3411 & 0.4872 & 0.3273 & 0.4721 & 207.47 \\
InfoLab.UDInfolab.RAG.bge.QueryAnsAI.tuned & RAG & 0.3646 & 0.4925 & 0.3394 & 0.4659 & 0.3264 & 0.4524 & 156.74 \\
softbank-meisei.ragtask-bm25-rank\_zephyr-gpt4o-llama70b & RAG & 0.3616 & 0.4830 & 0.3404 & 0.4603 & 0.3288 & 0.4480 & 226.60 \\
coordinators.baseline\_frag\_rag24.test\_command-r-plus\_top20 & RAG & 0.3596 & 0.4923 & 0.3535 & 0.4825 & 0.3496 & 0.4767 & 251.75 \\
WaterlooClarke.UWCrag\_stepbystep        & RAG & 0.3587 & 0.4883 & 0.3293 & 0.4570 & 0.3139 & 0.4406 & 157.99 \\
InfoLab.UDInfolab.AG-v2                  & AG & 0.3581 & 0.4947 & 0.3362 & 0.4688 & 0.3248 & 0.4555 & 158.92 \\
coordinators.baseline\_rag24.test\_command-r-plus\_top20 & AG & 0.3574 & 0.4942 & 0.3524 & 0.4850 & 0.3492 & 0.4796 & 251.75 \\
WaterlooClarke.UWCrag                    & RAG & 0.3571 & 0.4870 & 0.3324 & 0.4606 & 0.3193 & 0.4467 & 182.48 \\
webis.webis-ag-run0-taskrag              & AG & 0.3568 & 0.4941 & 0.3377 & 0.4736 & 0.3289 & 0.4639 & 157.44 \\
CIR.cir\_gpt-4o-mini\_Cosine\_20\_0.5\_100\_301\_p1 & AG & 0.3566 & 0.5020 & 0.3334 & 0.4779 & 0.3201 & 0.4644 & 148.83 \\
InfoLab.UDInfolab.RAG.AnsAI              & RAG & 0.3566 & 0.4889 & 0.3326 & 0.4616 & 0.3199 & 0.4473 & 157.95 \\
InfoLab.UDInfolab.AG-v1                  & AG & 0.3560 & 0.4943 & 0.3337 & 0.4684 & 0.3224 & 0.4550 & 159.54 \\
InfoLab.UDInfolab.bgeV2                  & RAG & 0.3553 & 0.4903 & 0.3296 & 0.4627 & 0.3152 & 0.4475 & 156.26 \\
KML.cohere+post\_processing              & AG & 0.3530 & 0.4936 & 0.3352 & 0.4747 & 0.3253 & 0.4647 & 330.48 \\
InfoLab.UDInfolab.RAG.Query              & RAG & 0.3511 & 0.4890 & 0.3270 & 0.4643 & 0.3142 & 0.4511 & 159.48 \\
h2oloo.listgalore\_l31-70b\_ragnarokv4\_top20 & RAG & 0.3493 & 0.4839 & 0.3239 & 0.4543 & 0.3106 & 0.4389 & 228.21 \\
CIR.cir\_gpt-4o-mini\_Jaccard\_50\_0.5\_100\_301\_p0 & AG & 0.3487 & 0.4967 & 0.3259 & 0.4712 & 0.3138 & 0.4577 & 139.48 \\
softbank-meisei.rag\_bm25-colbert\_faiss-gpt4o-llama70b & RAG & 0.3464 & 0.4745 & 0.3271 & 0.4539 & 0.3169 & 0.4429 & 189.36 \\
InfoLab.UDInfolab.RAG.bge.tuned          & RAG & 0.3455 & 0.4862 & 0.3236 & 0.4624 & 0.3121 & 0.4495 & 157.90 \\
InfoLab.UDInfolab.RAG.bge.QueryAgm.tuned & RAG & 0.3452 & 0.4849 & 0.3262 & 0.4622 & 0.3160 & 0.4500 & 159.37 \\
CIR.cir\_gpt-4o-mini\_Jaccard\_50\_1.0\_100\_301\_p0 & AG & 0.3434 & 0.4830 & 0.3191 & 0.4562 & 0.3060 & 0.4421 & 135.05 \\
coordinators.baseline\_rag24.test\_l31\_70b\_instruct\_top20 & AG & 0.3356 & 0.4643 & 0.3140 & 0.4392 & 0.3026 & 0.4261 & 196.79 \\
h2oloo.listgalore\_l31-70b\_ragnarokv4nocite\_top20 & RAG & 0.3243 & 0.4760 & 0.2967 & 0.4460 & 0.2813 & 0.4297 & 277.15 \\
citi.BEST\_gpt3.5                        & RAG & 0.3164 & 0.4398 & 0.2976 & 0.4197 & 0.2879 & 0.4090 & 183.21 \\
webis.webis-rag-run3-taskrag             & RAG & 0.3116 & 0.4247 & 0.2870 & 0.3976 & 0.2742 & 0.3836 & 153.80 \\
ielab.ielab-b70bf-70bqfs-ad\_hoc         & RAG & 0.3100 & 0.4390 & 0.2925 & 0.4191 & 0.2835 & 0.4086 & 205.42 \\
citi.SECOND\_cot\_gpt3.5                 & RAG & 0.3085 & 0.4336 & 0.2890 & 0.4108 & 0.2794 & 0.3992 & 191.21 \\
citi.BEST\_cot\_gpt3.5                   & RAG & 0.3077 & 0.4328 & 0.2933 & 0.4173 & 0.2854 & 0.4088 & 197.34 \\
Ruc01.ruc001                             & RAG & 0.3024 & 0.4258 & 0.2880 & 0.4086 & 0.2803 & 0.3996 & 237.08 \\
citi.SECOND\_gpt3.5                      & RAG & 0.2951 & 0.4288 & 0.2783 & 0.4083 & 0.2692 & 0.3975 & 179.55 \\
uog-tht.PG-mistral                       & RAG & 0.2906 & 0.4174 & 0.2772 & 0.4005 & 0.2690 & 0.3908 & 197.89 \\
ielab.ielab-b8bf-8bzs-ad\_hoc            & RAG & 0.2879 & 0.4184 & 0.2749 & 0.4045 & 0.2674 & 0.3968 & 175.12 \\
ielab.ielab-b8bf-8bfs-ad\_hoc            & RAG & 0.2849 & 0.4185 & 0.2724 & 0.4044 & 0.2650 & 0.3964 & 175.12 \\
citi.BEST\_gpt3.5\_another\_prompt       & RAG & 0.2820 & 0.4022 & 0.2642 & 0.3825 & 0.2546 & 0.3720 & 155.11 \\
IITD-IRL.zeph\_test\_rag\_rrf\_expand\_query & RAG & 0.2795 & 0.4024 & 0.2619 & 0.3809 & 0.2521 & 0.3691 & 87.31 \\
IITD-IRL.ag\_rag\_gpt35\_expansion\_rrf\_20 & AG & 0.2762 & 0.3992 & 0.2577 & 0.3761 & 0.2486 & 0.3645 & 90.60 \\
IITD-IRL.zeph\_test\_rag24\_doc\_query\_expansion+rrf & RAG & 0.2750 & 0.4028 & 0.2554 & 0.3801 & 0.2446 & 0.3678 & 88.50 \\
IITD-IRL.zeph\_test\_rag\_rrf\_raw\_query & RAG & 0.2739 & 0.4018 & 0.2562 & 0.3780 & 0.2463 & 0.3653 & 89.08 \\
citi.BEST\_gpt3.5\_new\_prompt           & RAG & 0.2712 & 0.3956 & 0.2539 & 0.3756 & 0.2444 & 0.3647 & 124.08 \\
buw.buw\_2                               & RAG & 0.2690 & 0.3994 & 0.2605 & 0.3904 & 0.2565 & 0.3858 & 218.23 \\
\multicolumn{9}{c}{\vdots} \\
\bottomrule
\end{tabular}}
\end{table*}
\begin{table*}
\scalebox{0.72}{
\begin{tabular}{lrccccccr}
\toprule
Run ID & Task & $V_{\textrm{strict}}$ &  $V$ & $W_{\textrm{strict}}$ & $W$ & $A_{\textrm{strict}}$ & $A$ & $L$ \\
\midrule
\multicolumn{9}{c}{\vdots} \\
 \\
IITD-IRL.ag\_rag\_gpt35\_expansion\_rrf\_15 & AG & 0.2677 & 0.3886 & 0.2500 & 0.3670 & 0.2415 & 0.3564 & 88.99 \\
buw.buw\_3                               & RAG & 0.2657 & 0.3954 & 0.2605 & 0.3882 & 0.2582 & 0.3844 & 218.23 \\
buw.buw\_5                               & RAG & 0.2641 & 0.3973 & 0.2568 & 0.3880 & 0.2527 & 0.3827 & 215.76 \\
TREMA-UNH.Enhanced\_Iterative\_Fact\_Refinement\ldots & AG & 0.2613 & 0.3961 & 0.2538 & 0.3857 & 0.2499 & 0.3802 & 375.42 \\
TREMA-UNH.Ranked\_Iterative\_Fact\_Extraction\ldots  & AG & 0.2605 & 0.3963 & 0.2523 & 0.3852 & 0.2485 & 0.3797 & 374.83 \\
citi.SECOND\_gpt3.5\_new\_prompt         & RAG & 0.2604 & 0.3802 & 0.2406 & 0.3581 & 0.2296 & 0.3462 & 123.03 \\
ielab.ielab-b8bf-8bp-8ba                 & RAG & 0.2591 & 0.3870 & 0.2402 & 0.3622 & 0.2306 & 0.3494 & 131.17 \\
IIIA-UNIPD.iiia\_standard\_p1\_reverse\_ht\_ag & AG & 0.2586 & 0.3786 & 0.2400 & 0.3570 & 0.2304 & 0.3460 & 141.68 \\
ielab.ielab-b8bf-8bp-8bafs               & RAG & 0.2582 & 0.3899 & 0.2396 & 0.3648 & 0.2300 & 0.3518 & 131.17 \\
buw.buw                                  & RAG & 0.2582 & 0.3902 & 0.2507 & 0.3806 & 0.2472 & 0.3760 & 197.46 \\
buw.oneshot\_post\_sentenced             & RAG & 0.2533 & 0.3772 & 0.2471 & 0.3701 & 0.2443 & 0.3665 & 243.98 \\
IITD-IRL.ag\_rag\_gpt35\_expansion\_rrf\_7 & AG & 0.2482 & 0.3674 & 0.2313 & 0.3472 & 0.2230 & 0.3371 & 86.55 \\
IIIA-UNIPD.iiia\_standard\_p1\_straight\_ag & AG & 0.2475 & 0.3680 & 0.2264 & 0.3443 & 0.2154 & 0.3320 & 135.92 \\
uis-iai.baseline\_top\_5                 & AG & 0.2471 & 0.3746 & 0.2296 & 0.3534 & 0.2197 & 0.3417 & 90.12 \\
webis.webis-rag-run4-reuserag            & RAG & 0.2462 & 0.3724 & 0.2302 & 0.3543 & 0.2209 & 0.3441 & 219.74 \\
IIIA-UNIPD.iiia\_standard\_p1\_straight\_ht\_ag & AG & 0.2461 & 0.3703 & 0.2280 & 0.3480 & 0.2186 & 0.3365 & 134.78 \\
IIIA-UNIPD.iiia\_standard\_p1\_reverse\_ag & AG & 0.2457 & 0.3669 & 0.2283 & 0.3457 & 0.2197 & 0.3351 & 139.48 \\
ielab.ielab-b8b-8bp-8bafs                & RAG & 0.2442 & 0.3610 & 0.2272 & 0.3416 & 0.2183 & 0.3313 & 114.82 \\
ielab.ielab-b70bf-70bqp-70bafs           & RAG & 0.2379 & 0.3599 & 0.2206 & 0.3358 & 0.2112 & 0.3229 & 85.65 \\
ielab.ielab-b-8bp-8bafs                  & RAG & 0.2340 & 0.3484 & 0.2197 & 0.3322 & 0.2119 & 0.3235 & 114.19 \\
uis-iai.ginger-fluency\_top\_5           & AG & 0.2128 & 0.3422 & 0.2044 & 0.3281 & 0.1996 & 0.3203 & 85.82 \\
uis-iai.ginger\_top\_5                   & AG & 0.2112 & 0.3398 & 0.2015 & 0.3263 & 0.1961 & 0.3190 & 96.11 \\
ielab.ielab-b70bf-70bqp-rarr             & RAG & 0.2038 & 0.3100 & 0.1885 & 0.2892 & 0.1801 & 0.2779 & 69.69 \\
webis.webis-ag-run2-reuserag             & AG & 0.2035 & 0.3287 & 0.1935 & 0.3176 & 0.1879 & 0.3112 & 225.54 \\
IITD-IRL.zeph\_test\_rag\_rrf\_expand\_query\_mistral & RAG & 0.2025 & 0.3219 & 0.1835 & 0.2976 & 0.1736 & 0.2850 & 70.89 \\
IITD-IRL.ag\_rag\_mistral\_expansion\_rrf\_20 & AG & 0.2017 & 0.3158 & 0.1839 & 0.2952 & 0.1751 & 0.2849 & 77.89 \\
IITD-IRL.zeph\_test\_rag\_rrf\_expand\_mistral\_top\_15 & RAG & 0.2000 & 0.3126 & 0.1855 & 0.2923 & 0.1772 & 0.2810 & 68.54 \\
IIIA-UNIPD.iiia\_standard\_p2\_straight\_ht\_ag & AG & 0.1993 & 0.3058 & 0.1811 & 0.2845 & 0.1717 & 0.2733 & 80.49 \\
IIIA-UNIPD.iiia\_standard\_p2\_straight\_ag & AG & 0.1984 & 0.3070 & 0.1774 & 0.2832 & 0.1666 & 0.2708 & 81.38 \\
IIIA-UNIPD.iiia\_standard\_p1\_straight  & RAG & 0.1973 & 0.3107 & 0.1813 & 0.2904 & 0.1726 & 0.2793 & 121.05 \\
coordinators.fs4\_bm25+rocchio\_snowael\_snowaem\ldots  & RAG & 0.1957 & 0.2965 & 0.1948 & 0.2938 & 0.1941 & 0.2924 & 208.32 \\
IIIA-UNIPD.iiia\_dedup\_p1\_reverse\_ht\_ag & AG & 0.1952 & 0.3155 & 0.1813 & 0.2966 & 0.1736 & 0.2865 & 104.35 \\
IIIA-UNIPD.iiia\_dedup\_p1\_reverse\_ag  & AG & 0.1949 & 0.3113 & 0.1769 & 0.2909 & 0.1672 & 0.2799 & 102.84 \\
uog-tht.FT-llama3                        & RAG & 0.1927 & 0.3211 & 0.1758 & 0.2986 & 0.1666 & 0.2866 & 79.11 \\
IITD-IRL.ag\_rag\_mistral\_expansion\_rrf\_15 & AG & 0.1925 & 0.3077 & 0.1756 & 0.2868 & 0.1670 & 0.2762 & 69.68 \\
uog-tht.ICL-mistral                      & RAG & 0.1920 & 0.3214 & 0.1747 & 0.2987 & 0.1651 & 0.2864 & 83.74 \\
IIIA-UNIPD.iiia\_dedup\_p1\_straight\_ag & AG & 0.1916 & 0.3100 & 0.1752 & 0.2885 & 0.1669 & 0.2776 & 100.95 \\
IIIA-UNIPD.iiia\_dedup\_p1\_straight\_ht\_ag & AG & 0.1914 & 0.3100 & 0.1754 & 0.2900 & 0.1671 & 0.2800 & 101.06 \\
IIIA-UNIPD.iiia\_standard\_p1\_reverse   & RAG & 0.1910 & 0.3034 & 0.1779 & 0.2883 & 0.1705 & 0.2800 & 122.51 \\
IIIA-UNIPD.iiia\_standard\_p2\_reverse\_ag & AG & 0.1897 & 0.2999 & 0.1744 & 0.2814 & 0.1665 & 0.2717 & 84.85 \\
IIIA-UNIPD.iiia\_standard\_p1\_reverse\_ht & RAG & 0.1865 & 0.3078 & 0.1720 & 0.2903 & 0.1643 & 0.2808 & 121.24 \\
IIIA-UNIPD.iiia\_standard\_p1\_straight\_ht & RAG & 0.1861 & 0.2985 & 0.1716 & 0.2822 & 0.1639 & 0.2733 & 121.21 \\
IIIA-UNIPD.iiia\_standard\_p2\_reverse\_ht\_ag & AG & 0.1855 & 0.3038 & 0.1712 & 0.2843 & 0.1638 & 0.2740 & 87.67 \\
IITD-IRL.zeph\_test\_rag\_rrf\_expand\_top\_5 & RAG & 0.1840 & 0.2965 & 0.1681 & 0.2756 & 0.1598 & 0.2647 & 64.78 \\
IITD-IRL.zeph\_test\_rag\_rrf\_expand\_top\_10 & RAG & 0.1815 & 0.2961 & 0.1715 & 0.2805 & 0.1656 & 0.2716 & 68.66 \\
IITD-IRL.ag\_rag\_mistral\_expansion\_rrf\_7 & AG & 0.1772 & 0.2886 & 0.1666 & 0.2757 & 0.1612 & 0.2691 & 70.94 \\
IRIT.ISIR-IRIT-zephyr\_query\_gen\_3p    & RAG & 0.1733 & 0.2988 & 0.1685 & 0.2877 & 0.1661 & 0.2819 & 170.26 \\
IRIT.ISIR-IRIT-zephyr\_sprompt\_3p       & RAG & 0.1655 & 0.2850 & 0.1543 & 0.2701 & 0.1490 & 0.2626 & 168.42 \\
IRIT.ISIR-IRIT-zephyr\_query\_gen        & RAG & 0.1623 & 0.2749 & 0.1575 & 0.2666 & 0.1544 & 0.2616 & 159.90 \\
IRIT.ISIR-IRIT-zephyr\_p2                & RAG & 0.1547 & 0.2650 & 0.1477 & 0.2539 & 0.1436 & 0.2476 & 153.05 \\
IIIA-UNIPD.iiia\_dedup\_p2\_straight\_ag & AG & 0.1539 & 0.2574 & 0.1373 & 0.2357 & 0.1286 & 0.2244 & 50.04 \\
IIIA-UNIPD.iiia\_dedup\_p2\_straight\_ht\_ag & AG & 0.1530 & 0.2558 & 0.1365 & 0.2347 & 0.1275 & 0.2234 & 48.93 \\
IIIA-UNIPD.iiia\_dedup\_p2\_reverse\_ht\_ag & AG & 0.1492 & 0.2579 & 0.1337 & 0.2368 & 0.1251 & 0.2255 & 53.65 \\
IIIA-UNIPD.iiia\_dedup\_p1\_reverse      & RAG & 0.1484 & 0.2519 & 0.1352 & 0.2356 & 0.1278 & 0.2268 & 88.40 \\
IIIA-UNIPD.iiia\_dedup\_p2\_reverse\_ag  & AG & 0.1470 & 0.2546 & 0.1313 & 0.2343 & 0.1225 & 0.2234 & 53.45 \\
IIIA-UNIPD.iiia\_dedup\_p1\_straight     & RAG & 0.1444 & 0.2526 & 0.1330 & 0.2369 & 0.1267 & 0.2285 & 86.54 \\
IIIA-UNIPD.iiia\_standard\_p2\_straight\_ht & RAG & 0.1403 & 0.2409 & 0.1288 & 0.2268 & 0.1227 & 0.2194 & 59.23 \\
webis.webis-rag-run5-reuserag            & RAG & 0.1399 & 0.2407 & 0.1337 & 0.2302 & 0.1296 & 0.2241 & 100.02 \\
ii\_research.iiresearch-bm25-top10-llama3-8b-instruct & RAG & 0.1385 & 0.2269 & 0.1301 & 0.2158 & 0.1250 & 0.2095 & 167.83 \\
IIIA-UNIPD.iiia\_standard\_p2\_straight  & RAG & 0.1381 & 0.2378 & 0.1282 & 0.2243 & 0.1227 & 0.2168 & 61.77 \\
IIIA-UNIPD.iiia\_standard\_p2\_reverse   & RAG & 0.1373 & 0.2276 & 0.1271 & 0.2148 & 0.1213 & 0.2077 & 61.78 \\
IIIA-UNIPD.iiia\_dedup\_p1\_straight\_ht & RAG & 0.1344 & 0.2401 & 0.1252 & 0.2278 & 0.1203 & 0.2212 & 85.74 \\
IIIA-UNIPD.iiia\_dedup\_p1\_reverse\_ht  & RAG & 0.1332 & 0.2389 & 0.1233 & 0.2253 & 0.1177 & 0.2177 & 86.98 \\
IIIA-UNIPD.iiia\_standard\_p2\_reverse\_ht & RAG & 0.1308 & 0.2226 & 0.1214 & 0.2110 & 0.1160 & 0.2045 & 61.12 \\
TREMA-UNH.Ranked\_Iterative\_Fact\ldots RIFER\ldots & RAG & 0.1276 & 0.2334 & 0.1225 & 0.2257 & 0.1199 & 0.2214 & 336.29 \\
SGU.qrant\_bge\_gemini                   & RAG & 0.1230 & 0.1871 & 0.1152 & 0.1789 & 0.1109 & 0.1746 & 123.78 \\
webis.webis-ag-run3-reuserag             & AG & 0.1184 & 0.2148 & 0.1138 & 0.2073 & 0.1114 & 0.2037 & 103.75 \\
IIIA-UNIPD.iiia\_dedup\_p2\_straight\_ht & RAG & 0.1091 & 0.1955 & 0.0979 & 0.1800 & 0.0923 & 0.1722 & 39.16 \\
IIIA-UNIPD.iiia\_dedup\_p2\_straight     & RAG & 0.1058 & 0.1950 & 0.0977 & 0.1822 & 0.0936 & 0.1757 & 38.82 \\
IIIA-UNIPD.iiia\_dedup\_p2\_reverse\_ht  & RAG & 0.1031 & 0.1860 & 0.0949 & 0.1746 & 0.0906 & 0.1683 & 40.01 \\
IIIA-UNIPD.iiia\_dedup\_p2\_reverse      & RAG & 0.1010 & 0.1915 & 0.0918 & 0.1758 & 0.0870 & 0.1674 & 38.83 \\
coordinators.anserini\_bm25.rag24.test\_top1 & RAG & 0.0703 & 0.1252 & 0.0681 & 0.1225 & 0.0671 & 0.1212 & 216.58 \\
webis.webis-manual                       & RAG & 0.0368 & 0.0510 & 0.0344 & 0.0478 & 0.0333 & 0.0464 & 19.44 \\
coordinators.test\_empty\_list           & RAG & 0.0000 & 0.0000 & 0.0000 & 0.0000 & 0.0000 & 0.0000 & 0.00 \\
\midrule
\textbf{Min}                               &  & 0.0000 & 0.0000 & 0.0000 & 0.0000 & 0.0000 & 0.0000 & 0.0000 \\
\textbf{Median}                            &  & 0.2673 & 0.4028 & 0.2508 & 0.3835 & 0.2416 & 0.3728 & 148.6728 \\
\textbf{Max}                               &  & 0.6549 & 0.7580 & 0.6077 & 0.7145 & 0.5927 & 0.7029 & 398.9867 \\
\bottomrule
\end{tabular}}
\vspace{0.25cm}
\caption{Scores for AG and RAG runs under the \autonuggets { / }\autoassign condition, over 301 topics that were provided to the participants.}
\label{tab:auto_nugget_auto_assignment_all_full}
\end{table*}

\section{Results}

For the TREC 2024 RAG Track, NIST received 93 runs from 20 groups for the RAG task and 53 runs from 11 groups for the AG task. 
Given resource constraints, the NIST annotators were able to evaluate only the two highest priority submissions from each group across the RAG and AG tasks, which translates into 31 runs from 18 groups for RAG and 14 runs from 9 groups for AG.

Note that human annotation is ongoing as of November 14, 2024.
The annotations completed by November 8, 2024 form the basis of partial results that we report here.
To be clear, our analyses are over 21 topics that have been fully judged across the 31 RAG runs and 14 AG runs discussed above.
We plan to provide more exhaustive analyses once annotations are complete.

Table~\ref{tab:edit_nugget_nist_assignment_all} shows scores for AG and RAG runs under the \autonuggetsedits { }/ \manulassign condition:\ automatic nugget creation with manual post-editing and manual nugget assignment.
Table~\ref{tab:auto_nugget_auto_assignment_all} shows scores for the same runs, over the same topics, but for the \autonuggets { } / \autoassign condition:\ automatic nugget creation and automatic nugget assignment.
These tables present results for (mostly) manual nugget evaluation and fully automatic nugget evaluation, respectively.

The biggest question, of course, is whether the manual scores and the automatic scores correlate?

The answer is, {\it yes}.
This is shown in Figure~\ref{fig:scatter_plots_1}, which is a scatter plot of $V_{\textrm{strict}}$ from Tables~\ref{tab:edit_nugget_nist_assignment_all} and~\ref{tab:auto_nugget_auto_assignment_all}.
The solid red circles and solid orange squares represent run-level $V_{\textrm{strict}}$ scores:\ red squares represent RAG runs and orange squares represent AG runs.
As reported in the top left corner, we observe a Kendall's $\tau$ correlation of 0.783, which represents substantial agreement, especially considering that we only have results over 21 topics.

Also in Figure~\ref{fig:scatter_plots_1}, the blue circles and purple squares show all topic/run combinations:\ blue circles for RAG and purple squares for AG.
If we treat each topic/run combination as an observation and compute Kendall's $\tau$ over all these observations, we arrive at a value of 0.324, which is rather low.
Alternatively, if we compute Kendall's $\tau$ for each topic, and then average across the per-topic correlations, we arrive at a figure of 0.518, also shown in the top-left corner.

The conclusion based on these results seem clear:\ while there is substantial disagreement on a topic-by-topic basis between the automatic and manual evaluation, at the run level, the rank correlation is substantial.
To repeat our top-level findings from the introduction:\
We observe a strong correlation between scores derived from a fully automatic nugget evaluation and a (mostly) manual nugget evaluation by NIST assessors.

Evaluation of RAG is, of course, complex and multi-faceted.
Our $V_{\textrm{strict}}$ score captures the presence of vital nuggets, but this needs to be balanced by the length of the answer.
Although our guidelines cap answer length at 400 words, there are teams who submitted answers that were much shorter on average.
Answer length is shown by $L$ in Table~\ref{tab:edit_nugget_nist_assignment_all} and Table~\ref{tab:auto_nugget_auto_assignment_all}, measured in terms of the total number of standard whitespace words in the answer.

To explore the impact of answer length on answer quality, Figure~\ref{fig:scatter_plots_5} plots $V_{\textrm{strict}}$ from Table~\ref{tab:edit_nugget_nist_assignment_all} (i.e., the \autonuggetsedits { }/ \manulassign condition) vs.\ length $L$, both at the run level and across all topic/run combinations.
We see that, indeed, there is a general correlation between $V_{\textrm{strict}}$ and answer length.
However, we also see relatively large variations in answer quality, even holding answer length constant:\ points that are vertically aligned have the same (mean) answer length, but contain different amounts of vital nuggets.
From this, we can trace a Pareto-optimal curve for the systems in the evaluation---that is, for a given answer length, what is the best that current systems can accomplish?

In these results, we have adopted the strategy of presenting RAG and AG submissions together.
In Figures~\ref{fig:scatter_plots_1} and~\ref{fig:scatter_plots_5}, this done with different colors and shapes.
While this approach increases clutter, it easily provides a comparison between RAG and AG runs as a whole.
Interesting, it does {\it not} appear to be the case that full RAG submissions outperform AG submissions in general, which suggests that the reference ranked lists that we provided to AG participants seem ``good enough'' from the perspective of answer generation.

Finally, in Table~\ref{tab:auto_nugget_auto_assignment_all_full}, we show scores under the \autonuggets { } / \autoassign condition for all runs submitted to evaluation, across all 301 topics provided to the participants.

Due to time constraints, we have not conducted all the analyses we wished to have performed.
The results and analyses that we present here answer many questions but raise even more, which we hope to address later.

\section{Conclusions}

While we are certainly not the first to propose a RAG evaluation methodology, we view our efforts as having two main distinguishing characteristics:\
First, by building on the nugget evaluation methodology dating back over two decades, we minimize reinvention of the wheel.
The information retrieval literature has a long tradition of deliberate and careful meta-evaluations that validate evaluation methodologies, and much work has examined different aspects of nugget evaluations.
For aspects of the evaluation that are not dependent on LLMs, we can simply build on existing findings.
Second, we have demonstrated strong correlations between scores derived from a fully automatic nugget evaluation and a (mostly) manual nugget evaluation.
Coupled with the first point, the manual scores for nugget evaluations have at least achieved reasonable acceptance by the community on how to evaluate free-form answers to complex questions.
We show that the evaluation methodology can be automated using LLMs.

Under the \autonuggets { } / \autoassign condition in our AutoNuggetizer framework, we are able to provide evaluation scores for all runs submitted to the TREC 2024 RAG Track, across all 301 topics---at only the cost of LLM inference.
Our framework has the potential to enable rapid iteration on RAG systems in a fully automatic manner, while providing some confidence that the automatically generated metrics have some degree of correlation to answer quality as determined by human assessors.
We can now potentially climb hills quickly and in a meaningful way!

\section*{Acknowledgments}

This work would not have been possible without Hoa Trang Dang, Ian Soboroff, and the team at NIST.
We are grateful for their contributions.
This research was supported in part by the Natural Sciences and Engineering Research Council (NSERC) of Canada.
Additional funding is provided by Snowflake and Microsoft via the Accelerating Foundation Models Research program.
Thanks to Corby Rosset for providing our Bing query test set, based on the methodology developed in~\citet{Rosset:2402.17896:2024}

\bibliographystyle{ACM-Reference-Format}
\bibliography{ref}


\begin{thebibliography}{18}


\ifx \showCODEN    \undefined \def \showCODEN     #1{\unskip}     \fi
\ifx \showDOI      \undefined \def \showDOI       #1{#1}\fi
\ifx \showISBNx    \undefined \def \showISBNx     #1{\unskip}     \fi
\ifx \showISBNxiii \undefined \def \showISBNxiii  #1{\unskip}     \fi
\ifx \showISSN     \undefined \def \showISSN      #1{\unskip}     \fi
\ifx \showLCCN     \undefined \def \showLCCN      #1{\unskip}     \fi
\ifx \shownote     \undefined \def \shownote      #1{#1}          \fi
\ifx \showarticletitle \undefined \def \showarticletitle #1{#1}   \fi
\ifx \showURL      \undefined \def \showURL       {\relax}        \fi
\providecommand\bibfield[2]{#2}
\providecommand\bibinfo[2]{#2}
\providecommand\natexlab[1]{#1}
\providecommand\showeprint[2][]{arXiv:#2}

\bibitem[Bajaj et~al\mbox{.}(2018)]%
        {Bajaj:1611.09268:2018}
\bibfield{author}{\bibinfo{person}{Payal Bajaj}, \bibinfo{person}{Daniel Campos}, \bibinfo{person}{Nick Craswell}, \bibinfo{person}{Li Deng}, \bibinfo{person}{Jianfeng Gao}, \bibinfo{person}{Xiaodong Liu}, \bibinfo{person}{Rangan Majumder}, \bibinfo{person}{Andrew McNamara}, \bibinfo{person}{Bhaskar Mitra}, \bibinfo{person}{Tri Nguyen}, \bibinfo{person}{Mir Rosenberg}, \bibinfo{person}{Xia Song}, \bibinfo{person}{Alina Stoica}, \bibinfo{person}{Saurabh Tiwary}, {and} \bibinfo{person}{Tong Wang}.} \bibinfo{year}{2018}\natexlab{}.
\newblock \showarticletitle{{MS} {MARCO}: A Human Generated MAchine Reading COmprehension Dataset}.
\newblock \bibinfo{journal}{\emph{arXiv:1611.09268}} (\bibinfo{year}{2018}).
\newblock


\bibitem[Craswell et~al\mbox{.}(2022)]%
        {craswell2022overview}
\bibfield{author}{\bibinfo{person}{Nick Craswell}, \bibinfo{person}{Bhaskar Mitra}, \bibinfo{person}{Emine Yilmaz}, \bibinfo{person}{Daniel Campos}, \bibinfo{person}{Jimmy Lin}, \bibinfo{person}{Ellen~M. Voorhees}, {and} \bibinfo{person}{Ian Soboroff}.} \bibinfo{year}{2022}\natexlab{}.
\newblock \showarticletitle{Overview of the {TREC} 2022 Deep Learning Track}. In \bibinfo{booktitle}{\emph{Proceedings of the Thirty-First Text REtrieval Conference (TREC 2022)}}. \bibinfo{address}{Gaithersburg, Maryland}.
\newblock


\bibitem[Dang and Lin(2007)]%
        {Dang_Lin_ACL2007}
\bibfield{author}{\bibinfo{person}{Hoa~Trang Dang} {and} \bibinfo{person}{Jimmy Lin}.} \bibinfo{year}{2007}\natexlab{}.
\newblock \showarticletitle{Different Structures for Evaluating Answers to Complex Questions:\ Pyramids Won't Topple, and Neither Will Human Assessors}. In \bibinfo{booktitle}{\emph{Proceedings of the 45th Annual Meeting of the Association for Computational Linguistics (ACL 2007)}}. \bibinfo{address}{Prague, Czech Republic}.
\newblock


\bibitem[Lassance et~al\mbox{.}(2023)]%
        {unoduolisto}
\bibfield{author}{\bibinfo{person}{Carlos Lassance}, \bibinfo{person}{Ronak Pradeep}, {and} \bibinfo{person}{Jimmy Lin}.} \bibinfo{year}{2023}\natexlab{}.
\newblock \showarticletitle{{Naverloo} @ {TREC} {Deep} {Learning} and {NeuCLIR} 2023: As Easy as Zero, One, Two, Three --- Cascading Dual Encoders, Mono, Duo, and Listo for Ad-Hoc Retrieval}. In \bibinfo{booktitle}{\emph{Proceedings of the Thirty-Second Text REtrieval Conference (TREC 2023)}}. \bibinfo{address}{Gaithersburg, Maryland}.
\newblock


\bibitem[Li et~al\mbox{.}(2023)]%
        {li2023towards}
\bibfield{author}{\bibinfo{person}{Zehan Li}, \bibinfo{person}{Xin Zhang}, \bibinfo{person}{Yanzhao Zhang}, \bibinfo{person}{Dingkun Long}, \bibinfo{person}{Pengjun Xie}, {and} \bibinfo{person}{Meishan Zhang}.} \bibinfo{year}{2023}\natexlab{}.
\newblock \showarticletitle{Towards general text embeddings with multi-stage contrastive learning}.
\newblock \bibinfo{journal}{\emph{arXiv:2308.03281}} (\bibinfo{year}{2023}).
\newblock


\bibitem[Lin and Demner-Fushman(2005)]%
        {Lin_Demner-Fushman_HLT-EMNLP2005}
\bibfield{author}{\bibinfo{person}{Jimmy Lin} {and} \bibinfo{person}{Dina Demner-Fushman}.} \bibinfo{year}{2005}\natexlab{}.
\newblock \showarticletitle{Automatically Evaluating Answers to Definition Questions}. In \bibinfo{booktitle}{\emph{Proceedings of the 2005 Human Language Technology Conference and Conference on Empirical Methods in Natural Language Processing (HLT/EMNLP 2005)}}. \bibinfo{address}{Vancouver, British Columbia, Canada}.
\newblock


\bibitem[Lin and Demner-Fushman(2006)]%
        {Lin_Demner-Fushman_IR2006}
\bibfield{author}{\bibinfo{person}{Jimmy Lin} {and} \bibinfo{person}{Dina Demner-Fushman}.} \bibinfo{year}{2006}\natexlab{}.
\newblock \showarticletitle{Methods for Automatically Evaluating Answers to Complex Questions}.
\newblock \bibinfo{journal}{\emph{Information Retrieval}} \bibinfo{volume}{9}, \bibinfo{number}{5} (\bibinfo{year}{2006}).
\newblock


\bibitem[Lin and Zhang(2007)]%
        {Lin_Zhang_SIGIR2007}
\bibfield{author}{\bibinfo{person}{Jimmy Lin} {and} \bibinfo{person}{Pengyi Zhang}.} \bibinfo{year}{2007}\natexlab{}.
\newblock \showarticletitle{Deconstructing Nuggets:\ The Stability and Reliability of Complex Question Answering Evaluation}. In \bibinfo{booktitle}{\emph{Proceedings of the 30th Annual International ACM SIGIR Conference on Research and Development in Information Retrieval (SIGIR 2007)}}. \bibinfo{address}{Amsterdam, the Netherlands}.
\newblock


\bibitem[Merrick et~al\mbox{.}(2024)]%
        {merrick:2024}
\bibfield{author}{\bibinfo{person}{Luke Merrick}, \bibinfo{person}{Danmei Xu}, \bibinfo{person}{Gaurav Nuti}, {and} \bibinfo{person}{Daniel Campos}.} \bibinfo{year}{2024}\natexlab{}.
\newblock \showarticletitle{Arctic-Embed: Scalable, Efficient, and Accurate Text Embedding Models}.
\newblock \bibinfo{journal}{\emph{arXiv:2405.05374}} (\bibinfo{year}{2024}).
\newblock


\bibitem[Nogueira et~al\mbox{.}(2020)]%
        {nogueira-etal-2020-document}
\bibfield{author}{\bibinfo{person}{Rodrigo Nogueira}, \bibinfo{person}{Zhiying Jiang}, \bibinfo{person}{Ronak Pradeep}, {and} \bibinfo{person}{Jimmy Lin}.} \bibinfo{year}{2020}\natexlab{}.
\newblock \showarticletitle{Document Ranking with a Pretrained Sequence-to-Sequence Model}. In \bibinfo{booktitle}{\emph{Findings of the Association for Computational Linguistics: EMNLP 2020}}. \bibinfo{publisher}{Association for Computational Linguistics}, \bibinfo{address}{Online}.
\newblock


\bibitem[Pradeep et~al\mbox{.}(2021)]%
        {EMD}
\bibfield{author}{\bibinfo{person}{Ronak Pradeep}, \bibinfo{person}{Rodrigo Nogueira}, {and} \bibinfo{person}{Jimmy Lin}.} \bibinfo{year}{2021}\natexlab{}.
\newblock \showarticletitle{The Expando-Mono-Duo Design Pattern for Text Ranking with Pretrained Sequence-to-Sequence Models}.
\newblock \bibinfo{journal}{\emph{arXiv:2101.05667}} (\bibinfo{year}{2021}).
\newblock


\bibitem[Pradeep et~al\mbox{.}(2023)]%
        {pradeep:2023b}
\bibfield{author}{\bibinfo{person}{Ronak Pradeep}, \bibinfo{person}{Sahel Sharifymoghaddam}, {and} \bibinfo{person}{Jimmy Lin}.} \bibinfo{year}{2023}\natexlab{}.
\newblock \showarticletitle{RankZephyr: Effective and Robust Zero-Shot Listwise Reranking is a Breeze!}
\newblock \bibinfo{journal}{\emph{CoRR}}  \bibinfo{volume}{abs/2312.02724} (\bibinfo{year}{2023}).
\newblock
\showeprint[arXiv]{2312.02724}


\bibitem[Pradeep et~al\mbox{.}(2024)]%
        {ragnarok}
\bibfield{author}{\bibinfo{person}{Ronak Pradeep}, \bibinfo{person}{Nandan Thakur}, \bibinfo{person}{Sahel Sharifymoghaddam}, \bibinfo{person}{Eric Zhang}, \bibinfo{person}{Ryan Nguyen}, \bibinfo{person}{Daniel Campos}, \bibinfo{person}{Nick Craswell}, {and} \bibinfo{person}{Jimmy Lin}.} \bibinfo{year}{2024}\natexlab{}.
\newblock \showarticletitle{{Ragnarök}: A Reusable RAG Framework and Baselines for TREC 2024 Retrieval-Augmented Generation Track}.
\newblock \bibinfo{journal}{\emph{arXiv:2406.16828}} (\bibinfo{year}{2024}).
\newblock


\bibitem[Rosset et~al\mbox{.}(2024)]%
        {Rosset:2402.17896:2024}
\bibfield{author}{\bibinfo{person}{Corby Rosset}, \bibinfo{person}{Ho-Lam Chung}, \bibinfo{person}{Guanghui Qin}, \bibinfo{person}{Ethan~C. Chau}, \bibinfo{person}{Zhuo Feng}, \bibinfo{person}{Ahmed Awadallah}, \bibinfo{person}{Jennifer Neville}, {and} \bibinfo{person}{Nikhil Rao}.} \bibinfo{year}{2024}\natexlab{}.
\newblock \showarticletitle{Researchy Questions: A Dataset of Multi-Perspective, Decompositional Questions for {LLM} Web Agents}.
\newblock \bibinfo{journal}{\emph{arXiv:2402.17896}} (\bibinfo{year}{2024}).
\newblock


\bibitem[Upadhyay et~al\mbox{.}(2024a)]%
        {Upadhyay:2411.08275:2024}
\bibfield{author}{\bibinfo{person}{Shivani Upadhyay}, \bibinfo{person}{Ronak Pradeep}, \bibinfo{person}{Nandan Thakur}, \bibinfo{person}{Daniel Campos}, \bibinfo{person}{Nick Craswell}, \bibinfo{person}{Ian Soboroff}, \bibinfo{person}{Hoa~Trang Dang}, {and} \bibinfo{person}{Jimmy Lin}.} \bibinfo{year}{2024}\natexlab{a}.
\newblock \showarticletitle{A Large-Scale Study of Relevance Assessments with Large Language Models: An Initial Look}.
\newblock \bibinfo{journal}{\emph{arXiv:2411.08275}} (\bibinfo{year}{2024}).
\newblock


\bibitem[Upadhyay et~al\mbox{.}(2024b)]%
        {Upadhyay_etal_arXiv2024}
\bibfield{author}{\bibinfo{person}{Shivani Upadhyay}, \bibinfo{person}{Ronak Pradeep}, \bibinfo{person}{Nandan Thakur}, \bibinfo{person}{Nick Craswell}, {and} \bibinfo{person}{Jimmy Lin}.} \bibinfo{year}{2024}\natexlab{b}.
\newblock \showarticletitle{{UMBRELA}: {UMbrela} is the (Open-Source Reproduction of the) {Bing} {RELevance} {Assessor}}.
\newblock \bibinfo{journal}{\emph{arXiv:2406.06519}} (\bibinfo{year}{2024}).
\newblock


\bibitem[Voorhees(2003)]%
        {Voorhees_TREC2003}
\bibfield{author}{\bibinfo{person}{Ellen~M. Voorhees}.} \bibinfo{year}{2003}\natexlab{}.
\newblock \showarticletitle{Overview of the {TREC} 2003 Question Answering Track}. In \bibinfo{booktitle}{\emph{Proceedings of the Twelfth Text REtrieval Conference (TREC 2003)}}. \bibinfo{address}{Gaithersburg, Maryland}.
\newblock


\bibitem[Yang et~al\mbox{.}(2017)]%
        {anserini}
\bibfield{author}{\bibinfo{person}{Peilin Yang}, \bibinfo{person}{Hui Fang}, {and} \bibinfo{person}{Jimmy Lin}.} \bibinfo{year}{2017}\natexlab{}.
\newblock \showarticletitle{{Anserini}: Enabling the Use of {Lucene} for Information Retrieval Research}. In \bibinfo{booktitle}{\emph{Proceedings of the 40th International ACM SIGIR Conference on Research and Development in Information Retrieval (SIGIR 2017)}}. \bibinfo{address}{Tokyo, Japan}.
\newblock


\end{thebibliography}

\end{document}